\documentclass[12pt]{article}
\usepackage{amsfonts,amsmath}
\usepackage[english]{babel}
\usepackage{amssymb}

\def\abz{\vspace{3mm} \noindent}

\newcommand{\al}{\alpha}
\newcommand{\gb}{\beta}
\newcommand{\gd}{\delta}
\newcommand{\gga}{\gamma}
\newcommand{\gl}{\lambda}
\newcommand{\gk}{\varkappa}
\newcommand{\gep}{\epsilon}
\newcommand{\dal}{\dot \alpha}
\newcommand{\dgb}{\dot \beta}
\newcommand{\dgga}{\dot \gamma}
\newcommand{\q}{\,,\qquad}
\newcommand{\p}{\partial}
\newcommand{\f}{\frac}
\newcommand{\fO}{\mathbf{\Omega}}
\newcommand{\fh}{\mathbf{h}}
\newcommand{\e}{\mathbf{q}}
\newcommand{\D}{\mathcal{D}}
\newcommand{\G}{\mathcal{G}}
\newcommand{\fD}{\mathcal{D}}
\newcommand{\fV}{\mathcal{V}}
\newcommand{\M}{\mathcal{M}}
\newcommand{\bM}{\mathbf{M}}
\newcommand{\bN}{\mathbf{N}}
\newcommand{\bR}{\mathbf{R}}
\newcommand{\E}{\mathcal{E}}
\newcommand{\T}{\mathbf{T}}
\newcommand{\g}{\mathrm{g}}
\newcommand{\F}{F}

\newcommand{\be}{\begin{equation}}
\newcommand{\ee}{\end{equation}}

\renewcommand{\kappa}{\varkappa}

\renewcommand{\theequation}{\thesection.\arabic{equation}}
\renewcommand{\thesection}{\arabic{section}}
\renewcommand{\thesubsection}{\arabic{section}.\arabic{subsection}}
\makeatletter \@addtoreset{equation}{section} \makeatother

\begin{document}

\begin{flushright}
FIAN/TD/01-09\\
\end{flushright}

\vspace{0.5cm}

\begin{center}
{\large\bf Unfolded Dynamics and
Parameter Flow of Generic $AdS_4$ Black Hole }

\vspace{1 cm}

{\bf V.E.~Didenko, A.S.~Matveev and M.A.~Vasiliev}\\
\vspace{0.5 cm} {\it
I.E. Tamm Department of Theoretical Physics, Lebedev Physical Institute,\\
Leninsky prospect 53, 119991, Moscow, Russia }

\vspace{0.6 cm}
didenko@lpi.ru, matveev@lpi.ru, vasiliev@lpi.ru \\
\end{center}

\vspace{0.4 cm}

\begin{abstract}
\noindent We present unfolded description of $AdS_4$  black hole
with generic parameters of mass, NUT,
magnetic and electric charges as well as two kinematical parameters
one of which is angular momentum.
A flow with respect to black hole parameters, that relates
 the obtained black hole unfolded system to the covariant
constancy condition for an $AdS_4$ global symmetry parameter, is found.
The proposed formulation gives rise to a coordinate-independent
description of the black hole metric in $AdS_4$. The
black hole charges are identified with flow evolution parameters
while its kinematical constants are the first integrals of the black
hole unfolded system expressed via invariants of the $AdS_4$ global
symmetry parameter. It is shown how the proposed method reproduces
various known forms of black hole metrics including the Carter  and
Kerr--Newman solutions.  Free flow gauge parameters allow us to
choose different metric representations such as Kerr--Schild,
double Kerr--Schild or generalized Carter--Plebanski in the
coordinate-independent way.
\end{abstract}

\newpage
\tableofcontents
\newpage

\section{Introduction}

In the  paper \cite{DMV} Kerr black hole in four-dimensional $AdS$
space-time was shown to admit unfolded formulation based on the
Killing equation and the equation for the so called Papapetrou
field. As a starting point we used the following well known facts
\cite{review}
\begin{itemize}
\item  Four-dimensional Einstein black holes are of Petrov D-type.
In asymptotically flat space Riemann tensor is
built of the derivatives of a Killing vector (Papapetrou field).
\item Kerr--Schild Ansatz reduces nonlinear
Einstein equations to linear Pauli--Fierz equations
both on flat and on $AdS$ background.
\end{itemize}
The first property was generalized in \cite{DMV} to the
black hole on $AdS_4$ space within spinor approach leading to the
description of $AdS_4$ Kerr black hole Weyl tensor in terms of
$AdS_4$ Papapetrou field. Then we were able to show that $AdS_4$
black hole Kerr--Schild vector has background covariant nature and
is built of $AdS_4$ Killing vector via certain coordinate-independent
Killing projectors. This allowed us to describe
Kerr black hole in $AdS_4$ covariant way via field redefinition of
the $AdS_4$ global symmetry parameter covariant constancy condition.
The important questions that have not been yet considered in
\cite{DMV} include
\begin{itemize}
\item  As the unfolded formulation is by construction coordinate-free
how does a particular choice of the $AdS_4$ global symmetry
parameter affects the diffeomorphic-invariant properties of the
resulting Kerr--Schild metric?
\item  How the approach of
\cite{DMV} can be generalized to a wider class of four-dimensional
black holes to include electric charge, NUT parameter, etc.?
\end{itemize}
In this paper we answer these and some related questions. In our
work we stick to the idea of \cite{DMV} that $AdS_4$ global
symmetry parameter covariant consistency equation
\be\label{D0}
D_0K_{AB}=0\q D_0^2=0\,,
\ee
where  $A, B=1,\dots, 4$ are the $AdS_4$ spinor indices and
$D_0$ is the $AdS_4$ covariant differential, admits a parametric
deformation into a wider class of black holes. A novelty,
however, is to rewrite \eqref{D0} in terms of Killing vector and
source-free Maxwell tensor field, rather than using Papapetrou
field as in \cite{DMV}. This redefinition turns out to be very
convenient being particularly natural taking
into account that all four-dimensional Einstein--Maxwell black
holes have curvature tensor built of a sourceless Maxwell tensor.
It follows that simple consistent deformation of \eqref{D0} that
preserves Killing and Maxwell properties of the system leads to
Petrov D-type Weyl tensor with Ricci tensor given by Maxwell
energy-momentum tensor and constant scalar curvature of $AdS_4$
space-time $(\Lambda=3\gl^2)$. Defined this way unfolded system
contains three real parameters $\M\in \mathbb{{C}}$ and $\e\in
\mathbb{R}$ instead of one real parameter of Kerr black
hole mass  of \cite{DMV}. We show that Re $\M$ and Im $\M$
correspond to black hole\footnote{The term black hole that
will be used throughout the paper is strictly  speaking abuse of
terminology as we do not restrict  metric parameters to the domain
that corresponds to the existence of horizons and absence of
naked singularities.} mass and NUT charge, respectively, while
$\e=2(e^2+\g^2)$, where $e$ and $\g$ are electric and magnetic
charges, respectively.

\abz In general, the obtained black hole unfolded system
reproduces the so called Carter--Plebanski family of solutions,
which in addition to aforementioned four real curvature parameters
(Re $\M$, Im $\M$, $\e$, $\gl$) have two kinematic constants
related to angular momentum $a$ and certain discrete parameter
$\gep$ \cite{Car,Carter4,Pleb}. These kinematic parameters are
shown to arise in the black hole unfolded system (BHUS) as two
invariants of unfolded equations (first integrals).

\abz To obtain explicit expressions for the metric resulting from
our unfolded equations and to validate it is indeed of
Carter--Plebanski family we use an efficient integrating flow
method analogous  to the one developed in \cite{PV,Gol} for higher
spin nonlinear equations. Applying the consistency requirement
$[\p_{\chi}, d]=0$ to BHUS, where $\chi=(\M, \e)$ are the
deformation parameters and $d$  is space-time de Rham
differential, we derive the first-order differential equations in
the $\chi$ parameter space for all fields involved, {\it i.e.,}
vierbein, Killing vector, etc. The obtained flow
equations can be easily integrated with the initial data $\M=0,
\e=0$ that correspond to pure $AdS_4$ vacuum leading to the
coordinate-free description of generic Carter--Plebanski family of
metrics.

\abz The integrating flow reveals remarkable properties of the
black hole parameters. In particular, the kinematic parameters
turn out to be related to two invariants
\be\label{Casimir}
C_2=\f14K_{AB}K^{AB}\q C_4=\f14\text{Tr}(K^4)
\ee
of the $AdS_4$ algebra $sp(4)$ which are modules that characterize
the vacuum unfolded system. For example, a static black hole
corresponds to
\be
C_4=C_2^2\,.
\ee

\abz One of the motivations for this work was to elaborate the
unfolded approach to classical black holes appropriate to the
analysis of black hole solutions in the spinor form of $4d$
higher spin gauge theory in the form of \cite{Ann,more,Gol}
(see also \cite{QG,sor,solv,SSS} for reviews of higher spin
theory). In  \cite{DMV} we have shown that at least at the free field
level the black hole solution admits a natural extension to higher spins.
The natural question for the future study is whether
those would receive corrections had the interaction switched on.
To study this interesting question it is necessary to have black
holes description in the spirit of higher spin unfolded
formulation of \cite{Ann,more,Gol}. In this paper we show that such
a formulation is indeed available.

\abz We believe, however, that the results of this paper may
on their own right have useful applications in black hole physics.
In particular, a wide class of black hole
metrics formulated in coordinate-free form allows one to
obtain straightforwardly their realization in any
background coordinates. Moreover, by choosing appropriately
free parameters in the integration flow one can
reduce metric to either Kerr--Schild or double Kerr--Schild or
``generalized" Carter--Plebanski form depending on the number of
deformation parameters.

\abz The rest of the paper is organized as follows. We start in
Section \ref{mainresults} by summarizing main results obtained in
this paper. In Section \ref{cart} we reformulate Einstein gravity
using  the Cartan formalism being most appropriate for our
analysis.  In Section \ref{AdS} starting from the Killing equation
in $AdS_4$ we rewrite it in the unfolded form. Then we study its
properties, particularly, find the first integrals, discrete
symmetry, introduce certain Killing projectors giving rise to four
Kerr--Schild vectors. In particular in Subsection \ref{AdS us_cas}
invariants of the $AdS_4$ symmetry algebra $sp(4)$ are discussed
in connection with Killing symmetries of the system. Generic black
hole unfolded system, obtained as a parametric deformation of
initial $AdS_4$ unfolded system,  is presented in Section
\ref{FDA}. We find that BHUS inherits most of the pre-deformed
properties and symmetries. In particular, it expresses Weyl tensor
in terms of Maxwell field making it manifestly of Petrov D-type.
In Section \ref{IF} we apply the integrating flow technique to
obtain the first order differential equations with respect to the
deformation parameters that encode generic black hole solution.
Integration of obtained flow-equations with $AdS_4$ initial data
is carried out in Section \ref{S-BHUS} giving rise to $AdS_4$
covariant and coordinate-independent description of black hole
metric. In Section \ref{coord-AdS} we use particular coordinate
system for $AdS_4$ space-time and its unfolded system. It allows
us to reproduce in Subsection \ref{coord-carter} canonical form of
Carter--Plebanski metric and to identify BHUS modules with the
physical black hole parameters. Section \ref{sum} contains summary
and conclusions.  The notation is summarized in Appendix
\ref{App-A}. The details on derivation of the integrating flow
equations and their integration are given in Appendices
\ref{App-B} and \ref{App-C}, respectively. For the readers'
convenience the unfolded equations are rewritten in vector
notation in Appendix \ref{App-D} and some useful properties are
presented in Appendix \ref{App-E}. Finally, Plebanski--Demianski
solution is commented in  Appendix \ref{App-F}.

\section{Main results} \label{mainresults}

The main result of our work is the unfolded formulation of generic
$AdS_4$ Einstein--Maxwell black hole solution. This formulation is
coordinate-independent. Modules of solutions include the real
dynamical modules $\bM$, $\bN$ and $\e$ which are, respectively,
the black hole mass, NUT and a combination of electric and
magnetic charges. The $o(3,2)\sim sp(4,\mathbb{\mathbb{R}})$
transformations act on the modules of BH solution including three
coordinates of the black hole position, three  Lorentz boosts
({\it i.e.,} velocities) and two angles of the rotation axis
orientation. Two invariants of the $AdS_4$ transformations
parameterize black hole kinematical parameters -- its angular
momentum per unit mass $a$ and Carter--Plebanski parameter $\gep$.
The normalization of the $AdS_4$ invariants with $\gep=\pm1$ or 0
sets the scale for black hole curvature modules $\bM$, $\bN$ and
$\e$. The charge $\e=2(e^2+\g^2)$ arises as some inner
$u(1)$-invariant, while the electro-magnetic duality mixes
$\bM\leftrightarrow\bN$ and $e\leftrightarrow\g$.

\abz To reproduce generic black hole we use the idea of \cite{DMV},
constructing black hole unfolded system as a deformation
of $AdS_4$ global symmetry constancy equation
\be
D_0K_{AB} = 0\,, \label{ads-const}
\ee
where $K_{AB}(x)=K_{BA}(x)$ is an $AdS_4$ symmetry parameter and
$D_0$  is the $AdS_4$ covariant differential (for notation see
Appendix \ref{App-A}). As explained in \cite{Vas} any solution of
(\ref{ads-const}) describes some symmetry of $AdS_4$. In
particular, it gives rise to the corresponding Killing vector (see
e.g. \cite{solv}).

\abz Indeed, in two-component spinor notation $K_{AB}$ has the
form
\be \label{ads-kappa1}
K_{AB} = \left(
\begin{array}{cc}
\lambda^{-1}\kappa_{\alpha\beta} & V_{\alpha\dot\beta} \\
V_{\beta\dot\alpha} & \lambda^{-1}
\bar\kappa_{\dot\alpha\dot\beta}
\end{array}
\right),
\ee
where $\lambda$ is the $AdS$ radius and  $V_{\alpha\dot\alpha}$ is
some vector. From \eqref{ads-kappa1} it follows that
$V_{\al\dal}$ satisfies
\be\label{kil-spin}
DV_{\al\dal}=\f12h^{\gga}{}_{\dal}\gk_{\gga\al}+\f12h_{\al}{}^{\dgga}\bar\gk_{\dal\dgga}\,,
\ee
where $D$ is the Lorentz derivative and $h_{\al\dal}$ is the
$AdS_4$ vierbein one-form. From \eqref{kil-spin} it follows
immediately that
\be
D_{\al\dal}V_{\al\dal}=0\,,
\ee
or, equivalently, in tensor notation
\be\label{kil}
D_i V_j+D_j V_i = 0\,.
\ee
The equation \eqref{kil} means that $V_{\al\dal}$ is a Killing
vector. \eqref{ads-const} does not impose any additional
conditions being equivalent to \eqref{kil} along with the $AdS_4$
consistency \eqref{D0}. The fields $\kappa_{\alpha\alpha}$ and
$\bar\kappa_{\dot\alpha\dot\alpha}$ are (anti)self-dual parts of
the Killing two-form\footnote{One can easily check that it is a
closed Killing--Yano tensor. Note that the reverse statement
generally is not true, {\it i.e.,} vector associated with closed
Killing--Yano tensor is not necessary a Killing vector.} $\kappa_{ij}=D_iV_j$
($i,j=1,\ldots,4$ are world indices).

\abz Note that the system \eqref{ads-const} written down in components
\eqref{ads-kappa1} provides the simplest example of unfolded
equations consistent by virtue of zero-curvature condition
(\ref{D0}) for $AdS_4$. In Section \ref{FDA} we will show that a
simple consistent deformation of \eqref{ads-const} leads to
certain Killing--Maxwell unfolded system that describes generic
Carter--Plebanski metric. In turns out that it can be written down
in terms of the fields of \eqref{ads-kappa1} in a $AdS_4$
covariant way.

\abz To reproduce the metric explicitly we show that
\eqref{ads-const} generates four Kerr--Schild vectors built of
components \eqref{ads-kappa1} in a coordinate-independent way.
Two of them, $k^i$ and  $n^i$, are real
\be
k_{i}k^{i}=n_{i}n^{i}=0\q k^{i}D_{i}k_{j}=n^{i}D_{i}n_{j}=0
\ee
and another two are complex-conjugated and orthogonal to $k^{i}$
and $n^i$
\be
l^{-+}_{i}l^{-+i}=l^{+-}_{i}l^{+-i}=0\q
l^{-+i}D_{i}l^{-+}_{j}=l^{+-i}D_{i}l^{+-}_{j}=0\,.
\ee
Their explicit realization in terms of $AdS_4$ fields $K_{AB}$
will be given in Section \ref{AdS}. To write down black hole
metric in $AdS_4$ covariant and coordinate-free form we introduce
the following Lorentz scalars
\be\label{G}
\G=\frac{\lambda^2}{\sqrt{-\kappa^2}}\q
\bar{\G}=\frac{\lambda^2}{\sqrt{-\bar{\kappa}^2}}\,,
\ee
where\footnote{For notations used throughout this paper see
Appendix \ref{App-A}.}
$\kappa_{\alpha\beta}\kappa^{\beta}{}_{\gamma}=\kappa^2
\varepsilon_{\alpha\gamma}$.  This allows us to define ``canonical
scalars''\footnote{The reason of this name is that in a certain
reference frame, scalars $r$ and $y$ are equal to the canonical
coordinates introduced by Carter in \cite{Carter4}.}
\be\label{can-coord}
2r = \frac{1}{\G}+\frac{1}{\bar\G}, \quad 2iy = \frac{1}{\bar\G}-
\frac{1}{\G}.
\ee

\abz Using the unfolded analysis along with the integration flow
method we show that the solution of the obtained first order flow
equations describes generic Einstein--Maxwell black hole on
$AdS_4$ space-time in coordinate-independent form
\begin{align}\label{ds-gen}
ds^2 &= ds_0^2 + \frac{2\bM r-\frac{\e}{2}}{r^2+y^2} (\alpha_1(r)
k_idx^i+ \alpha_2(r) n_idx^i)^2 -
\frac{2\bN y+\frac{\e}{2}}{r^2+y^2} (\gb_1(y) l^{+-}_idx^i+ \gb_2(y) l^{-+}_idx^i)^2\notag \\
&+
4\alpha_1(r)\alpha_2(r)\frac{r^2+y^2}{\Delta_r\hat\Delta_r}(2\bM
r-\frac{\e}{2})dr^2 -
4\gb_1(y)\gb_2(y)\frac{r^2+y^2}{\Delta_y\hat\Delta_y}(2\bN
y+\frac{\e}{2})dy^2,
\end{align}
where $\alpha_1(r), \alpha_2(r)$ and $\gb_1(y), \gb_2(y)$
 subjected to the  constraints
\be \alpha_1(r)+\alpha_2(r)= 1, \qquad
\gb_1(y)+\gb_2(y)=1,
\ee
are otherwise arbitrary, parameterizing some  gauge ambiguity,
$ds^2_0$ is the background $AdS_4$ metric,
$\hat\Delta_r $ and $\hat\Delta_y$ are the following polynomials
\begin{eqnarray}
\hat\Delta_r &=& 2\mathbf{M}r + r^2(\lambda^2r^2+I_1) + \frac12(-\e+\frac{I_2}{2}),  \\
\hat\Delta_y &=& 2\mathbf{N}y + y^2(\lambda^2y^2-I_1) +
\frac12(\e+\frac{I_2}{2}),
\end{eqnarray}
and
\be
\Delta_{r} =  \left.\hat\Delta_{r}\right|_{\bM,\bN,\e =0}=
r^2(\lambda^2r^2+I_1) + \frac14I_2\,,
\ee
\be
\Delta_{y} =  \left.\hat\Delta_{y}\right|_{\bM,\bN,\e =0}=
y^2(\lambda^2y^2-I_1) + \frac14I_2\,,
\ee
with $I_1, I_2$ being some first integrals of \eqref{ads-const}
related to the invariants \eqref{Casimir} as follows
\be
C_2=I_1\q C_4=I^2_1+\gl^2I_2\,.
\ee

\abz
Note that, generally, the metric \eqref{ds-gen} is complex.
Reality of the metric requires
\be
\gb_1=\gb_2=\frac12\,.
\ee
However, sometimes, it may be useful to consider complex metrics
(for example, to reproduce the double Kerr--Schild form
\cite{Pleb}).

\abz Black hole Maxwell field $F=dA$ is generated by a one-form potential
that, up to a gauge freedom, can be chosen in the form
\be
A=\frac{r}{r^2+y^2} k_{i}dx^{i}.
\ee

\abz The metric \eqref{ds-gen} is valid for any values of its
parameters. However, in the case of zero NUT parameter $\bN=0$ the
flow integration can be performed  differently  giving rise to a
simpler expression for the black hole metric. In, particular we
will show how the familiar Kerr--Schild form for Kerr--Newman
black hole \cite{Newman} can be obtained in arbitrary coordinates.

\abz The solution \eqref{ds-gen} is characterized by two
polynomial functions whose coefficients are determined by six
arbitrary parameters. It belongs to Petrov D-type \cite{Petrov}
class of solutions of Einstein--Maxwell equations including
non-zero cosmological constant and electro-magnetic field such
that the two degenerate principal null congruences of the Weyl
tensor are aligned with the two principal null congruences of the
Maxwell tensor. $\bM$ plays the role of mass, $\bN$ is a NUT
charge, $\lambda^2$ is the cosmological term, $I_2$ is a
rotational parameter $a$, and $I_1$ is the Carter--Plebanski
parameter which can be set equal to 1, 0 or -1 by a rescaling
transformation discussed below. It will be shown that
$\e=2(e^2+\g^2)$,  where $e$ and $\g$ are electric and magnetic
charges respectively. Note, that the charges enter \eqref{ds-gen}
via $\e$ combination and thus can not be distinguished unless some
external charged fields introduced.

\abz Particular solution types depend on the values of the
curvature parameters -- $\bM, \bN, \e$ and $sp(4)$ invariants. Let
us enlist the main cases:

\begin{itemize}
\item Carter--Plebanski solution
\end{itemize}

\abz All of the six parameters are non-zero. The metric is given by
\eqref{ds-gen}. It is easy to write it down in the well-known
Carter--Plebanski form \cite{Carter4, Pleb} setting
$\al_1=\al_2=\gb_1=\gb_2=\frac12$ (see Subsection
\ref{coord-carter}). The rotational parameter is
 $a^2=I_2/4$, whereas the Carter--Plebanski parameter is
$\gep=I_1$.

\begin{itemize}
\item Double Kerr--Schild form of Carter--Plebanski
\end{itemize}

\abz The gauge choice $\alpha_1=\gb_1=1$, $\alpha_2=\gb_2=0$ leads to
the so called double Kerr--Schild form of \eqref{ds-gen}
\be
ds^2 = ds_0^2 +  \frac{2r}{r^2+y^2}\left(\mathbf{M} -
\frac{\e}{4r}\right)k_ik_jdx^idx^j -
\frac{2y}{r^2+y^2}\left(\mathbf{N} + \frac{\e}{4y}\right)
 l^{+-}_il^{+-}_jdx^idx^j
\ee
which is complex in Minkowski signature.

\abz The following cases of zero NUT charge are important for
physical applications:

\begin{itemize}
\item $\bN=0$, $C_2=1+\gl^2 a^2$, $C_4=C_2^2+4\gl^2 a^2$
\end{itemize}

\abz This case provides Kerr--Newman  solution with $a$ being
black hole angular momentum per unit mass. The metric can be
written down in the Kerr--Schild form \cite{K-S}
\be
ds^2 = ds_0^2 + \frac{2\bM r-\e}{r^2+y^2}k_ik_jdx^idx^j\,.
\ee

\begin{itemize}
\item $\bN=0$, $C_4=C_2^2$ (equivalently, $K_{A}{}^{C}K_{C}{}^{B}=C_2\gd_{A}{}^{B})$
\end{itemize}

\abz The particular case of non-rotating solution results from the
further degeneration $I_2 =0$, $y=0$. It describes
Reissner--Nordstr\"{o}m solution. Again, one can conveniently put
the metric in Kerr--Schild form
\be
ds^2 = ds_0^2 +\left( \frac{2\bM}{r}-\frac{\e}{r^2}\right)
k_ik_jdx^idx^j\,.
\ee

\abz Let us note, that all listed solutions are invariant under
the rescaling
\be
\label{scaling}
K_{AB}\to \mu K_{AB}\,,\quad \bM\to\mu^3\bM\,,\quad
\bN\to\mu^3\bN\,,\quad \e^2\to\mu^4\e^2
\ee
with real constant $\mu$ that yields
\be
C_2\to\mu^2C_2\q C_4\to\mu^4C_4\,.
\ee
This means that among two kinematical parameters of \eqref{ds-gen}
one can be always chosen to be discrete 1, 0 or -1. Alternatively,
using the scaling ambiguity (\ref{scaling}) one can scale away the
mass parameter $\bM$ that will then be represented by the
parameter $\epsilon$.

\abz In the following we will essentially use two-component spinor language
which has great advantages in $4d$ description, so let us proceed
to Cartan formalism of gravity.

\section{Cartan formalism} \label{cart}

In Riemannian approach to black hole in $AdS_4$ gravity the metric
and Maxwell gauge field verify Einstein--Maxwell equations
\be \label{einst}
R_{ij} = 3\lambda^2 g_{ij} + T_{ij}\,,
\ee
\be
D_iF^{i}{}_{j}=0\,
\ee
with the energy-momentum tensor of the form
\be  \label{Tij}
T_{ij} = 4(e^2+\g^2) \left(F_{ik}F_j{}^k -\frac14 g_{ij}F_{kl}F^{kl}\right).
\ee
Let us proceed to Cartan formulation of gravity. Let $dx^m
\fO_{m}{}^{ab}$ be an antisymmetric Lorentz connection one-form
and $dx^m \fh_m{}^a$ be a vierbein one-form. These can be
identified with the  gauge fields of the $AdS_4$ symmetry algebra
$o(3,2)$. The corresponding $AdS_4$ curvatures $\bR^{ab}=\frac12
\bR_{ij}{}^{ab}dx^i\wedge dx^j$ and $\bR^a=\frac12
\bR_{ij}{}^adx^i\wedge dx^j$ have the form
\be \label{1Cartan}
\bR^{ab} = d\fO^{ab} + \fO^{ac}\wedge\fO_{c}{}^{b}
-\lambda^2 \fh^a \wedge \fh^b\,,
\ee
\be \label{2Cartan}
\bR^a = d\fh^a + \fO^{ac}\wedge \fh_c\,,
\ee
where  $a,b,c=0,\dots, 3$ are Lorentz indices. Lorentz indices are
raised and lowered with the flat metric $\eta_{ab}
=\text{diag}(1,-1,-1,-1)$. The zero-torsion condition $\mathbf{R}^a =0$
expresses algebraically the Lorentz connection $\fO$ via
derivatives of $\fh$. Then the $\lambda$-independent part of the
curvature two-form \eqref{1Cartan} identifies with the Riemann
tensor.

\abz For the case of non-zero energy-momentum tensor it is
convenient to decompose the curvature two-form into its traceless
part associated with the Weyl tensor and tracefull one provided by
$T_{ij}$
\be \label{CWM}
\mathbf{R}_{ab} = \frac12 \fh^c \wedge \fh^d
C_{cdab}+\frac12(\fh_a\T_b-\fh_b\T_a)\,,
\ee
where $C_{abcd}$ is the Weyl tensor in the local frame,
$C_{abcd}=-C_{bacd}=-C_{abdc}=C_{cdab}$ and $\T_a=T_{ab}\fh^b$ is
a one-form associated with the energy-momentum tensor. Equation
\eqref{CWM} is equivalent to the metric form of Einstein equations
\eqref{einst} with the metric
\be
g_{mn} =\fh_m{}^a \fh_n{}^b\eta_{ab}\,.
\ee

\abz Now we proceed to spinor reformulation of Einstein--Maxwell
theory. Einstein equation \eqref{CWM} and torsion-free condition
\eqref{2Cartan} can be rewritten in the spinor notation as
follows\footnote{See Appendix \ref{App-A} for notation.}. Lorentz
connection one-forms $\fO_{\alpha\alpha},
\bar\fO_{\dot\alpha\dot\alpha}$ and vierbein one-form
$\fh_{\alpha\dot\alpha}$ can be identified with the gauge fields
of $sp(4) \sim o(3,2)$.  It is easy to check that the equivalent
spinor form of \eqref{Tij} is
\be
T_{\alpha\dot\alpha\beta\dot\beta} = -4(e^2+\g^2)F_{\alpha\beta} \bar{F}_{\dot\alpha\dot\beta}.
\ee
It is obviously invariant under the electro-magnetic duality transformation
\be \label{se-m}
F_{\alpha\alpha} \to e^{i\theta}F_{\alpha\alpha}, \quad
\bar{F}_{\dot\alpha\dot\alpha} \to
e^{-i\theta}\bar{F}_{\dot\alpha\dot\alpha}.
\ee
Then Einstein equations with
cosmological constant acquire the form\footnote{The symmetrization
over denoted by the same letter spinor indices is implied.}
\be \label{RHC}
\mathcal{R}_{\alpha\alpha} =  d\fO_{\alpha\alpha} +
\frac12\fO_{\alpha}{}^\gamma \wedge \fO_{\gamma\alpha} =
\frac{\lambda^2}{2}\, \mathbf{H}_{\alpha\alpha} +
\frac18 \mathbf{H}^{\gamma\gamma}C_{\gamma\gamma\alpha\alpha}
+ \frac{e^2+\g^2}{2} \bar{\mathbf{H}}^{\dot\gamma\dot\gamma}\bar{F}_{\dot\gamma\dot\gamma}F_{\alpha\alpha}
\ee
\be \label{RHCc}
\mathcal{\bar R}_{\dot\alpha\dot\alpha} =  d\bar{\fO}_{\dot\alpha\dot\alpha} +
\frac12 \bar{\fO}_{\dot\alpha}{}^{\dot\gamma} \wedge \bar{\fO}_{\dot\gamma\dot\alpha}=
\frac{\lambda^2}{2}\, \mathbf{\bar{H}}_{\dot\alpha\dot\alpha}  +
\frac18 \mathbf{\bar H}^{\dot\gamma\dot\gamma}\bar{C}_{\dot\gamma\dot\gamma\dot\alpha\dot\alpha}
+ \frac{e^2+\g^2}{2} \mathbf{H}^{\gamma\gamma} F_{\gamma\gamma}\bar{F}_{\dot\alpha\dot\alpha}
\ee
\be \label{Tfree}
\mathcal{R}_{\alpha\dot{\alpha}} = d\fh_{\alpha\dot{\alpha}} +
\frac12 \fO_{\alpha}{}^{\gamma}\wedge \fh_{\gamma\dot{\alpha}} +
\frac12 \bar{\fO}_{\dot{\alpha}}{}^{\dot{\gamma}}\wedge
\fh_{\alpha\dot{\gamma}}=0\,,
\ee
where $\mathcal{R}_{\alpha\beta}$ and
$\bar{\mathcal{R}}_{\dot\alpha\dot\beta}$ are the components of the Loretnz curvature two-form
\be \label{D2Rcal}
\fD^2\xi_{\alpha\dot\alpha} = \frac12 \mathcal{R}_{\alpha}{}^{\beta}\xi_{\beta\dot\alpha}
+ \frac12 \mathcal{\bar R}_{\dot\alpha}{}^{\dot\beta}\xi_{\alpha\dot\beta}
\ee
and
\be \label{Hcap}
\mathbf{H}^{\alpha\alpha} = \fh^{\alpha}{} _{\dot\alpha} \wedge
\fh^{\alpha\dot\alpha}\,, \qquad
\bar{\mathbf{H}}^{\dot\alpha\dot\alpha} =
\fh_{\alpha}{}^{\dot\alpha} \wedge \fh^{\alpha\dot\alpha}\,.
\ee

\section{$AdS_4$ unfolded system}\label{AdS}

\subsection{Killing equations unfolded}\label{AdS us}

Let us formulate the unfolded system that describes $AdS_4$ geometry
along with some its global symmetry. We start with an $AdS_4$ Killing vector
$V^m$ and its covariant derivative
\be\label{k2f}
\kappa_{mn} = D_m V_n, \quad \kappa_{mn} = -\kappa_{nm}\,,
\ee
which will be referred to as the Killing two-form or Papapetrou
field.

\abz Since the $AdS_4$ Riemann curvature has vanishing Weyl tensor
one can write down the following system
\be\label{FDA1-ads}
DV_{\alpha\dot\alpha}  = \frac12
h^{\gamma}{}_{\dot\alpha}\kappa_{\gamma\alpha}+\f12h_{\al}{}^{\dot\gamma}\bar{\gk}_{\dal\dgga}\,,
\ee
\be\label{FDA2-ads}
D\kappa_{\alpha\alpha} =\lambda^2
h_{\alpha}{}^{\dot\gamma}V_{\alpha\dot\gamma}\,,
\ee
\be\label{FDA2c-ads}
D\bar\kappa_{\dot\alpha\dot\alpha} =\lambda^2
h^{\gamma}{}_{\dot\alpha} V_{\gamma\dot\alpha}\,,
\ee
which is consistent provided that
\be\label{FDA23}
Dh_{\alpha\dot\alpha}=0\,,
\ee
\be\label{FDA3-ads}
R_{\alpha\alpha} \equiv d\Omega_{\alpha\alpha} +
\frac12\Omega_{\alpha}{}^{\beta} \wedge \Omega_{\beta\alpha} =
\f{\lambda^2}{2}h_{\alpha\dot\alpha} \wedge
h_{\alpha}{}^{\dot\alpha}\,,
\ee
\be\label{FDA4-ads}
\bar R_{\dot\alpha\dot\alpha} \equiv d\bar\Omega_{\dot\alpha\dot\alpha}
+ \frac12\bar\Omega_{\dot\alpha}{}^{\dot\beta} \wedge
\bar\Omega_{\dot\beta\dot\alpha} =
 \f{\lambda^2}{2}h_{\alpha\dot \alpha}
\wedge h^{\alpha}{}_{\dot\alpha}\,,
\ee
where  $h_{\al\dal}$ is the $AdS_4$ vierbein,
$\Omega_{\alpha\alpha}$ and $\bar\Omega_{\dot\alpha\dot\alpha}$
are components of Lorentz connection, $D$ is the background
Lorentz covariant differential and $R_{\alpha\alpha}$, $\bar
R_{\dot\alpha\dot\alpha}$ are the components of $AdS_4$ curvature
two-form
$$
D^2\xi_{\alpha\dot\alpha}=\frac12 R_{\alpha}{}^{\beta}\xi_{\beta\dot\alpha}
+ \frac12 \bar R_{\dot\alpha}{}^{\dot\beta}\xi_{\alpha\dot\beta}.
$$

\abz The equations \eqref{FDA1-ads}--\eqref{FDA4-ads} can be
rewritten in the manifestly $AdS_4$ covariant form. Indeed, let
$K_{AB}$ be the $AdS_4$ zero-form
\be \label{ads-kappa}
K_{AB} = \left(
\begin{array}{cc}
\lambda^{-1}\kappa_{\alpha\beta} & V_{\alpha\dot\beta} \\
V_{\beta\dot\alpha} & \lambda^{-1}
\bar\kappa_{\dot\alpha\dot\beta}
\end{array}
\right),
\ee
and $\Omega_{AB}$ be the frame one-form
\be
\Omega_{AB} = \left(
\begin{array}{cc}
\Omega_{\alpha\beta} & -\lambda h_{\alpha\dot\beta} \\
-\lambda h_{\beta\dot\alpha} & \bar\Omega_{\dot\alpha\dot\beta}
\end{array}
\right)\,.
\ee
Then the manifestly $sp(4)$ covariant form of the
 system \eqref{FDA1-ads}--\eqref{FDA4-ads} reads as
\begin{eqnarray}
D_0K_{AB} & =& 0, \label{ads const1} \\
R_{0AB} &\equiv& d\Omega_{AB} + \frac12\Omega_{A}{}^{C} \wedge
\Omega_{CB} = 0, \label{ads const2}
\end{eqnarray}
where $D_0$ is the $AdS_4$ covariant differential. The first
equation is the covariant constancy condition for global symmetry
parameter, while the second one describes $AdS_4$.

\abz Let us note, that the system \eqref{FDA1-ads}--\eqref{FDA2c-ads} was used
in \cite{DMV} as a starting point in the construction of the
Kerr black hole unfolded system. The deformation was
performed in terms of Killing vector $V_{\al\dal}$ and Papapetrou
field $\gk_{\al\al}$, $\overline\gk_{\dal\dal}$. However, it turns
out more convenient to rescale the Papapetrou field appropriately,
rewriting the $AdS_4$ unfolded equations using the rescaled field.
So, let us introduce self-dual Maxwell tensor $F_{\al\al}$ and
its complex conjugate $\bar{F}_{\dal\dal}$ as
\be \label{F-vac}
F_{\alpha\alpha}=-\lambda^{-2}\G^3 \kappa_{\alpha\alpha}\q
\bar{F}_{\dal\dal}=-\gl^{-2}\bar{\G}^3\bar{\gk}_{\dal\dal}\,,
\ee
where
\be \label{G-ads}
\G=\frac{\lambda^2}{\sqrt{-\kappa^2}} = (-F^2)^{1/4}\q
\bar{\G}=\frac{\lambda^2}{\sqrt{-\bar{\kappa}^2}} =
(-\bar{F}^2)^{1/4}
\ee
and the roots  on the right hand sides of \eqref{G-ads}
are chosen so as to have $\G$ and $\overline\G$ complex conjugated.

\abz Then \eqref{FDA1-ads}--\eqref{FDA2c-ads} can be rewritten
as\footnote{Parentheses mean symmetrization over indices.}
\begin{eqnarray}
DV_{\alpha\dot\alpha} & =& \frac12\rho\,
h^{\gamma}{}_{\dot\alpha}F_{\gamma\alpha}+\frac12 \bar{\rho}\, h_{\al}{}^{\dot\gamma}\bar{F}_{\dal\dgga}\,, \label{vac1-ads} \\
DF_{\alpha\alpha} &=&-\frac{3}{2\G}
h^{\beta\dot\gamma}V^{\beta}{}_{\dot\gamma}F_{(\beta\beta}F_{\alpha\alpha)}\,, \label{vac2-ads} \\
D\bar{F}_{\dot\alpha\dot\alpha} &=&-\frac{3}{2\bar{\G}}
h^{\gamma\dot\beta}V_{\gamma}{}^{\dot\beta}\bar{F}_{(\dot\beta\dot\beta}\bar{F}_{\dot\alpha\dot\alpha)}\,.\label{vac2c-ads}
\end{eqnarray}
with
\be \label{rho-ads}
\rho=-\lambda^2 \G^{-3}, \qquad \bar\rho=-\lambda^2 \bar\G^{-3}.
\ee
In what follows this Killing--Maxwell system\footnote{The first
step towards the analysis of Killing--Maxwell system was made by
Carter in \cite{Carter3}, where the relation between sourceless
Maxwell field and Killing--Yano tensor was discussed.} along with the
$AdS_4$ curvature equations \eqref{FDA23}--\eqref{FDA4-ads} will
be referred to as $AdS_4$ unfolded system. Note, that so defined
field strength \eqref{F-vac} is well defined in the flat limit
$\gl\to 0$.

\abz The important property of \eqref{vac1-ads}--\eqref{vac2c-ads}
is that $F_{\al\al}$ and $\bar F_{\dal\dal}$ satisfy source-free
Maxwell equations and Bianchi identities
\be \label{max+bian}
D_{\gamma\dot\alpha}F_{\alpha}{}^{\gamma} = 0\,, \qquad
D_{\alpha\dot\gamma}\bar{F}_{\dot\alpha}{}^{\dot\gamma} = 0.
\ee
Using \eqref{vac2-ads} and \eqref{vac2c-ads} one obtains useful
equations for $\G$ and $\bar\G$
\be \label{dG}
d\G=-\frac12 h^{\alpha\dot\alpha}V^{\alpha}{}_{\dot\alpha}F_{\alpha\alpha}, \qquad
d\bar\G=-\frac12 h^{\alpha\dot\alpha}V_{\alpha}{}^{\dot\alpha}\bar{F}_{\dot\alpha\dot\alpha}\,.
\ee
Unfolded equations \eqref{vac1-ads}--\eqref{vac2c-ads} have a
number of remarkable properties. In particular, the system can be
shown to possess Killing--Yano tensor and another Killing vector
built of $V_{\al\dal}$ and $F_{\al\al}$, $\bar{F}_{\dal\dal}$.
These properties are summarized in Appendix \ref{App-E}.

\abz An important property related to the description
of the kinematical parameters of $4d$ black holes is that the system
\eqref{vac1-ads}--\eqref{vac2c-ads} possesses two first integrals
\begin{eqnarray}\label{ints}
I_1 &=& V^2 - \frac{\lambda^2}{2}\left(\frac{1}{\G^2}+\frac{1}{\bar\G^2}\right), \label{vac-FI1} \\
I_2 &=& \frac{1}{\G^3\bar{\G}^3}V^{\alpha\dot\alpha}V^{\alpha\dot\alpha}F_{\alpha\alpha}
\bar{F}_{\dot\alpha\dot\alpha}  - V^2 \left(\frac{1}{\G^2}+\frac{1}{\bar\G^2}\right)+
\frac{\lambda^2}{4}\left(\frac{1}{\G^2}-\frac{1}{\bar\G^2}\right)^2\,,\qquad \label{vac-FI2}
\end{eqnarray}
where $V^2 = \frac12 V_{\alpha\dot\alpha}V^{\alpha\dot\alpha}$.
Using \eqref{vac1-ads}--\eqref{vac2c-ads} one can straightforwardly
check that  $dI_1=0$ and $dI_2=0$. Obviously enough, these
conserved quantities are related to two invariants of $sp(4)$
algebra as we will see more explicitly in Subsection \ref{AdS
us_cas}.

\abz Finally, $AdS_4$ unfolded system is invariant under the
following transformation
\be \label{bhussym} \tau_\mu:
\quad (V_{\alpha\dot\alpha}, \F_{\alpha\alpha},
\bar\F_{\dot\alpha\dot\alpha}) \rightarrow (\mu
V_{\alpha\dot\alpha}, \frac{1}{\mu|\mu|}\F_{\alpha\alpha},
\frac{1}{\mu|\mu|}\bar\F_{\dot\alpha\dot\alpha}),
\ee where $\mu$ is a real parameter. Yet another symmetry of the system is the parity
transformation
\be\label{parity}
\pi: (V_{\alpha\dot\alpha}, h_{\alpha\dot\alpha})
 \rightarrow
(-V_{\alpha\dot\alpha}, -h_{\alpha\dot\alpha}).
\ee

\subsection{Killing projectors} \label{S:KP}

As explained in \cite{DMV}, the key element of the black hole
unfolding, that eventually gives rise to \eqref{ds-gen}, is the
construction of Kerr--Schild vectors out of the $AdS_4$ global
symmetry parameter. The proposed procedure is essentially
four-dimensional being based on certain projectors we are in a position
to define, namely, we split the spinor space into two orthogonal
sectors using the projectors constructed from the Maxwell field.
In what follows they will be referred  to as Killing projectors.

\abz Let two pairs of mutually conjugated projectors
$\Pi^{\pm}_{\al\gb}$ and $\bar\Pi^{\pm}_{\dal\dgb}$ have the form
\be \label{introP}
\Pi^\pm_{\alpha\beta} = \frac12 (\epsilon_{\alpha\beta}  \pm
\frac{1}{\G^2}\F_{\alpha\beta}), \qquad
\bar\Pi^\pm_{\dot\alpha\dot\beta} = \frac12
(\epsilon_{\dot\alpha\dot\beta}  \pm
\frac{1}{\bar\G^2}\bar\F_{\dot\alpha\dot\beta})\,,
\ee
so that
\be
\Pi^+_{\alpha\beta}+\Pi^-_{\alpha\beta}=\gep_{\al\gb}\q
\bar\Pi^+_{\dal\dgb}+\bar\Pi^-_{\dal\dgb}=\gep_{\dal\dgb}\,,
\ee
and
\be
\Pi^\pm_{\alpha}{}^{\beta}\Pi^\pm_{\beta\gamma} =
\Pi^\pm_{\alpha\gamma}\,, \qquad
\Pi^\pm_{\alpha}{}^{\beta}\Pi^\mp_{\beta\gamma} = 0\q
\bar\Pi^\pm_{\dot\alpha}{}^{\dot\beta}\bar\Pi^\pm_{\dot\beta\dot\gamma}
= \bar\Pi^\pm_{\dot\alpha\dot\gamma}\,, \qquad
\bar\Pi^\pm_{\dot\alpha}{}^{\dot\beta}\bar\Pi^\mp_{\dot\beta\dot\gamma}
= 0\,.
\ee
From the definition \eqref{introP} it follows
\be
\Pi^\pm_{\alpha\beta} = - \Pi^\mp_{\beta\alpha}\,, \qquad
\bar\Pi^\pm_{\dot\alpha\dot\beta} = -
\bar\Pi^\mp_{\dot\beta\dot\alpha}\,.
\ee
From \eqref{introP}, \eqref{vac2-ads} and \eqref{dG} one finds the
following differential properties
\begin{eqnarray}
D\Pi^\pm_{\alpha\beta} &=& \pm \f{\G}{2}
(\Pi^{+}_{\alpha\gamma}\Pi^{+}_{\beta\gamma}+\Pi^{-}_{\alpha\gamma}\Pi^{-}_{\beta\gamma})
h^{\gamma}{}_{\dot\gamma}V^{\gamma\dot\gamma},  \label{derPi} \\
D\bar\Pi^\pm_{\dot\alpha\dot\beta} &=& \pm \f{\bar\G}{2}
(\bar\Pi^{+}_{\dot\alpha\dot\gamma}\bar\Pi^{+}_{\dot\beta\dot\gamma}+\bar\Pi^{-}_{\dot\alpha\dot\gamma}\bar\Pi^{-}_{\dot\beta\dot\gamma})
h_{\gamma}{}^{\dot\gamma}V^{\gamma\dot\gamma}.  \label{derPic}
\end{eqnarray}
Hereinafter we will focus on the holomorphic ({\it i.e.},
undotted) sector of the system. All relations in the antiholomorphic
sector result by conjugation.

\abz The projectors \eqref{introP} split the two-dimensional
(anti)holomorphic spinor space into the direct sum of two
one-dimensional subspaces. For any $\psi_\alpha$ we set
\be
\psi^\pm_{\alpha} = \Pi^\pm_{\alpha}{}^{\beta} \psi_{\beta}\,,
\qquad \psi^+_\alpha+\psi^-_\alpha = \psi_\alpha,
\ee
so that
$
\Pi^\mp_{\alpha}{}^{\beta} \psi^\pm_{\beta} = 0\,.
$
This allows us to build light-like vectors with the aid of
the projectors. Indeed, consider an arbitrary vector
$V_{\alpha\dot\alpha}$. Using \eqref{introP} define
$V^{\pm}_{\alpha\dot\alpha}$ and
$V^{\pm\mp}_{\alpha\dot\alpha}$ as
\be\label{def-V+-}
V^{\pm}_{\alpha\dot\alpha} = \Pi^\pm_{\alpha}{}^{\beta}
\bar\Pi^\pm_{\dot\alpha}{}^{\dot\beta} V_{\beta\dot\beta}\q
V^{+-}_{\alpha\dot\alpha} =  \Pi^+_{\alpha}{}^{\beta}
\bar\Pi^-_{\dot\alpha}{}^{\dot\beta} V_{\beta\dot\beta}\,, \qquad
V^{-+}_{\alpha\dot\alpha} =  \Pi^-_{\alpha}{}^{\beta}
\bar\Pi^+_{\dot\alpha}{}^{\dot\beta} V_{\beta\dot\beta}\,.
\ee
Since the projectors have rank one, they can be expressed
via a pair of some basis spinors $(\xi_\al,
\eta_\al)$ as follows
\be \label{Pi-xieta}
\Pi^+_{\alpha\beta} =
\frac{\eta_{\alpha}\xi_{\beta}}{\eta_\gamma\xi^{\gamma}}, \qquad
\Pi^-_{\alpha\beta} =
\frac{\xi_{\alpha}\eta_{\beta}}{\xi_\gamma\eta^{\gamma}}.
\ee
\abz Obviously, $V^{\pm}_{\alpha\dot\beta}V^{\pm\alpha\dot\gamma}
= 0$ and $V^{\pm}_{\alpha\dot\alpha}V^{\pm\beta\dot\alpha} = 0$.
Then $V^{\pm}_{\alpha\dot\alpha}$ and
$V^{\pm\mp}_{\alpha\dot\alpha}$ can be cast into the form
\be \label{Vxieta}
V^{-}_{\alpha\dot\alpha} = \xi_{\alpha} \bar\xi_{\dot\alpha}, \quad
V^{+}_{\alpha\dot\alpha} = \eta_{\alpha} \bar\eta_{\dot\alpha}, \quad
V^{+-}_{\alpha\dot\alpha} = q\,\eta_{\alpha} \bar\xi_{\dot\alpha}, \quad
V^{-+}_{\alpha\dot\alpha} = \bar{q}\,\xi_{\alpha} \bar\eta_{\dot\alpha},
\ee
where $q(x)$ and $\bar q(x)$ are some complex functions. As a
consequence of \eqref{Pi-xieta}, we also have
\be \label{F-xieta}
\F_{\alpha\alpha} = 2\G^2 \frac{\xi_{\alpha}\eta_{\alpha}}{\eta_\gamma\xi^{\gamma}}.
\ee
Now, from \eqref{Vxieta} it is obvious that
\be
V^{\pm}_{\alpha\dot\beta}V^{\pm}_{\beta\dot\alpha} =
V^{\pm}_{\alpha\dot\alpha}V^{\pm}_{\beta\dot\beta}\,, \qquad
V^{-+}_{\alpha\dot\beta}V^{+-}_{\beta\dot\alpha} =
-\frac{(V^{-+}V^{+-})}{(V^{-}V^+)}
V^{-}_{\alpha\dot\alpha}V^{+}_{\beta\dot\beta}\,,
\ee
where
$$
(AB) = A_{\alpha\dot\alpha}B^{\alpha\dot\alpha}.
$$
Also note that
\be
V^2 = (V^{+}V^{-}) + (V^{+-}V^{-+}) = (1-q\bar{q})(V^{+}V^{-}).
\ee
It is  worth to note that according to Papapetrou \cite{Papa} any
stationary axisymmetric solution of empty-space Einstein's
equations have a discrete symmetry upon simultaneous inversion of
the angular and time Killing vectors. Boyer and Lindquist
\cite{BL} have written a special transformation which casts the
empty-space Kerr metric into a form manifestly invariant under
such an inversion. In the $AdS$ unfolded system, this symmetry
is  $\tau_{-1}$ \eqref{bhussym} that interchanges the projectors
\be \label{tau-1}
\tau_{-1}: \quad \Pi^\pm_{\alpha\beta} \to \Pi^\mp_{\alpha\beta},
\quad  \bar\Pi^\pm_{\dot\alpha\dot\beta} \to \bar\Pi^\mp_{\dot\alpha\dot\beta}.
\ee

\subsection{Kerr--Schild null-vector basis} \label{S:ngc}

Now we are in a position to introduce the complete set of null-vectors
(complex null tetrad)
\begin{eqnarray}
k_{\alpha\dot\alpha} &=&
\frac{2}{(V^{+}V^-)}V^{-}_{\alpha\dot\alpha}, \qquad
n_{\alpha\dot\alpha} = \frac{2}{(V^{+}V^-)}V^{+}_{\alpha\dot\alpha},  \label{kn-def} \\
l^{+-}_{\alpha\dot\alpha} &=&
\frac{2}{(V^{+-}V^{-+})}V^{+-}_{\alpha\dot\alpha}, \qquad
l^{-+}_{\alpha\dot\alpha} =
\frac{2}{(V^{+-}V^{-+})}V^{-+}_{\alpha\dot\alpha}. \label{ll-def}
\end{eqnarray}
Note, that $k_{\al\dal}$ and $n_{\al\dal}$ are real vectors,
whereas $l^{+-}_{\al\dal}$ and $l^{-+}_{\al\dal}$ are mutually
conjugated
\[
l^{+-}_{\al\dal}=\bar{l}^{-+}_{\al\dal}\,.
\]
The discrete  symmetry $\tau_{-1}$ \eqref{tau-1} interchanges the
null-vectors
\begin{eqnarray}
\tau_{-1}&:& k_{\alpha\dot\alpha} \to n_{\alpha\dot\alpha}, \\
\tau_{-1}&:& l^{+-}_{\alpha\dot\alpha} \to l^{-+}_{\alpha\dot\alpha} .
\end{eqnarray}
Schematically, in terms of spinors \eqref{Pi-xieta},
 one can think of these null
vectors as
\be\label{e-decomp}
k_{\al\dal}\sim\xi_{\al}\bar{\xi}_{\dal}\,,\quad
n_{\al\dal}\sim\eta_{\al}\bar{\eta}_{\dal}\,,\quad
l^{+-}_{\al\dal}\sim\eta_{\al}\bar{\xi}_{\dal}\,,\quad
l^{-+}_{\al\dal}\sim\xi_{\al}\bar{\eta}_{\dal}\,.
\ee

\abz It is convenient to arrange this set of null-vectors into the array
\be
e_{I, \alpha\dot\alpha} = \left(  k_{\alpha\dot\alpha},  n_{\alpha\dot\alpha},
l^{+-}_{\alpha\dot\alpha},  l^{-+}_{\alpha\dot\alpha}  \right)   \label{eA-def}
\ee
with the evident properties
\be
e_{I,\alpha\dot\alpha} e_{I,}{}^{\alpha\dot\alpha} = 0, \qquad \f12 e_{I,\alpha\dot\alpha}V^{\alpha\dot\alpha}=1,
\ee
where $I=1,\ldots,4$ (no summation over $I$).
Obviously,
\be
(e_1\,e_2)= \frac{4}{(V^+V^-)}, \qquad
(e_3\,e_4 )= \frac{4}{(V^{+-}V^{-+})}.
\ee
{}From \eqref{vac1-ads}--\eqref{vac2c-ads} it follows that
\begin{eqnarray} \label{DeA}
D e_{I,\alpha\dot\alpha}  &=&
(-1)^{\sigma_I}\frac{\G}{4} \left( \rho\,\G e_{I,\alpha\dot\alpha}e_{I,\beta\dot\beta}
+ e_{I,\beta}{}^{\dot\gamma}V_{\alpha\dot\gamma} e_{I,}{}^{\gamma}{}_{\dot\alpha}V_{\gamma\dot\beta}  \right)h^{\beta\dot\beta} \notag \\
&+& (-1)^{\bar\sigma_I}\frac{\bar\G}{4} \left( \bar\rho\,\bar\G e_{I,\alpha\dot\alpha}e_{I,\beta\dot\beta}
+ e_{I,\alpha}{}^{\dot\gamma}V_{\beta\dot\gamma} e_{I,}{}^{\gamma}{}_{\dot\beta}V_{\gamma\dot\alpha}  \right)h^{\beta\dot\beta},
\end{eqnarray}
where $\sigma_I$ counts the number of $\eta_{\alpha}$ in
$e_{I,\alpha\dot\alpha}$ \eqref{eA-def}, {\it i.e.,} $\sigma_I =
(0,1,1,0)$ and $\bar\sigma_I$ counts the number of
$\bar\eta_{\dal}$, {\it i.e.,} $\bar\sigma_I= (0,1,0,1)$.

\abz A simple consequence of \eqref{DeA} and \eqref{e-decomp} is
that $e_{I,\alpha\dot\alpha}$ obey the geodesity condition
\be \label{congr}
 e_{I,}{}^{\alpha\dot\alpha}D_{\alpha\dot\alpha} e_{I,\beta\dot\beta}=0\,.
\ee
(No summation over $I$). In other words, all null-vectors
\eqref{kn-def} and \eqref{ll-def} are Kerr--Schild, that is
light-like and each satisfying \eqref{congr}. In addition, $e_{I,
\alpha\dot\alpha}$ are eigenvectors of the Maxwell tensors
$\F_{\alpha\alpha}$, $\bar\F_{\dot\alpha\dot\alpha}$ as follows
from \eqref{kn-def}, \eqref{ll-def} and \eqref{Vxieta}
\begin{eqnarray}
\F_{\alpha\beta}e_{I,}{}^{\beta}{}_{\dot\alpha} &=& (-1)^{\sigma_I}\G^2 e_{I,\alpha\dot\alpha},\label{fege} \\
\bar\F_{\dot\alpha\dot\beta}e_{I,\alpha}{}^{\dot\beta} &=&
(-1)^{\bar\sigma_I}\bar\G^2 e_{I,\alpha\dot\alpha}.\label{fegec}
\end{eqnarray}
For the tensor version of these and related formulae we refer the
reader to Appendix \ref{App-D}.

\abz As a consequence of \eqref{vac1-ads}--\eqref{vac2c-ads} and
\eqref{dG} the following properties can be verified
\be\label{prop1}
D_{\alpha\dot\alpha} e_{I,}{}^{\alpha\dot\alpha}=-2((-1)^{\sigma_I}\G+(-1)^{\bar\sigma_I}\bar{\G})\,, \qquad
e_{I,}{}^{\alpha\dot\alpha}D_{\alpha\dot\alpha}\G=2(-1)^{\sigma_I}\G^2\,,
\ee
\be\label{prop2}
D_{\alpha\dot\alpha}
(((-1)^{\bar\sigma_I}\G+(-1)^{\sigma_I}\bar{\G})e_{I,}{}^{\alpha\dot\alpha})
= -4 \G\bar{\G}\,,\qquad
D_{\alpha\dot\alpha} (\G\bar{\G}e_{I,}{}^{\alpha\dot\alpha})=0\,,
\ee
\be \label{prop3}
e_{I,\alpha}{}^{\dot\alpha} D_{\alpha\dot\alpha} e_{I,\gamma\dot\gamma}
= (-1)^{\sigma_I}\G  e_{I\alpha}{}^{\dot\alpha} V_{\gamma\dot\alpha} e_{I,\alpha\dot\gamma}.
\ee
Using \eqref{dG} and \eqref{DeA} one can make sure that each
Kerr--Schild vector \eqref{eA-def} generates the Maxwell field
\eqref{F-vac} via
\be \label{max-dif}
F_{\alpha\alpha}=\frac12 D_{\alpha\dot\alpha}
((\G+(-1)^{\sigma_I+\bar\sigma_I}\bar{\G})e_{I,\alpha}{}^{\dot\alpha}),
\ee
\be \label{maxc-dif}
\bar{F}_{\dot\alpha\dot\alpha}=\frac12 D_{\alpha\dot\alpha}
(( \bar\G + (-1)^{\sigma_I+\bar\sigma_I}\G)e_{I,}{}^{\alpha}{}_{\dot\alpha})\,.
\ee

\abz Now let us  give the explicit expressions for $(V^+V^-)$ and
$(V^{+-}V^{-+})$ which will be useful in what follows. Using
\eqref{vac-FI1} and \eqref{vac-FI2} and making change of variables
\eqref{can-coord} we obtain
\begin{eqnarray}
(V^+V^-) &=& \frac{\Delta_r}{r^2+y^2}, \label{vac-vpp}\\
(V^{+-}V^{-+}) &=& -\frac{\Delta_y}{r^2+y^2}, \label{vac-vpm}
\end{eqnarray}
with
\begin{eqnarray}
\Delta_r &=&  r^2(\lambda^2r^2+I_1) + \frac{I_2}{4}\,, \label{vac-Deltar} \\
\Delta_y &=& y^2(\lambda^2y^2-I_1) +
\frac{I_2}{4}\,.\label{vac-Deltay}
\end{eqnarray}

\abz The important remark is that this way we define the so called
Carter canonical coordinates  $r$ and $y$ (see \cite{Frolov2} for
more detail) which naturally  arise in our approach, being related
to the Maxwell field.

\abz Let us introduce one-forms $\E_I $ corresponding to the
null-vectors \eqref{eA-def} which will play an important role in
metric construction
\be
\E_I =\frac12  e_{I,\alpha\dot\alpha} h^{\alpha\dot\alpha}=(K,N,
L^{+-},L^{-+})\,. \label{vac-eform-def}
\ee
Using \eqref{max-dif} and \eqref{maxc-dif} we observe that
the vector-potentials
\be \label{vac-A12}
A_{1,2}=\frac{r}{r^2+y^2}\E_{1,2}
\ee
generate the same Maxwell tensor field $F=dA_{1,2}$.

\abz The second pair of vector-potentials
\be \label{vac-A34}
A_{3,4}=\frac{y}{r^2+y^2}\E_{3,4}
\ee
gives the Hodge dual field strength $*F=dA_{3,4}$.

\abz One can check that
\be \label{vac e1-e2}
K-N = \frac{2(r^2+y^2)}{\Delta_r}dr, \qquad L^{+-}-L^{-+} =
\frac{2(r^2+y^2)}{i\Delta_y}dy.
\ee
From here it is obvious  that the one-form potentials $A_{1,2}$
\eqref{vac-A12} ( $A_{3,4}$  \eqref{vac-A34})
belong to the same gauge class and generate the same
Maxwell field $F_{ij}$ ($*F_{ij}$).

\subsection{$AdS_4$ invariants}\label{AdS us_cas}

To reveal the algebraic nature of the first integrals \eqref{ints}
it is instructive to use the $AdS_4$ covariant form \eqref{ads
const1}--\eqref{ads const2} of the unfolded system
\eqref{FDA1-ads}--\eqref{FDA4-ads}.

\abz Consider the $AdS_4$ invariants constructed out of $K_{AB}$.
Calculation of the square of $K_{AB}$ yields
\be \label{kap^2}
K_{AC}K^C{}_B  = \left(
\begin{array}{cc}
(V^2+\lambda^{-2}\kappa^2)\varepsilon_{\alpha\beta} &
\gl^{-1}(\kappa_{\alpha\gamma}V^{\gamma}{}_{\dot\beta} -
\bar\kappa_{\dot\gamma\dot\beta}V_{\alpha}{}^{\dot\gamma})
\\ -\gl^{-1}(\kappa_{\beta\gamma}V^{\gamma}{}_{\dot\alpha} -
\bar\kappa_{\dot\gamma\dot\alpha}V_{\beta}{}^{\dot\gamma})
& (V^2+\lambda^{-2}\bar\kappa^2)\varepsilon_{\dot\alpha\dot\beta}
\end{array}
\right).
\ee
$AdS_4$ indices are raised and lowered with the aid of canonical
$sp(4)$-form (see Appendix \ref{App-A}). All higher powers of
$K_{AB}$ have the same structure with the scalar coefficients
changed. The two independent $sp(4)$ invariants are
\begin{eqnarray}
C_2 &=& \frac14 K_{AB}K^{AB}  = I_1, \label{ads-c1} \\
C_4 &=& \frac14 \text{Tr}(K^4) = I_1^2 + \lambda^2 I_2,
\label{ads-c2}
\end{eqnarray}
where $I_{1,2}$ are defined in \eqref{vac-FI1} and \eqref{vac-FI2}.
Note that  all odd invariants are  zero
$
\frac14 \text{Tr}(K^n) = 0\,,
$
for odd $n$.
All higher even invariants are expressed in terms of $C_{2,4}$
(equivalently, $I_{1,2}$) in the agreement with the fact
that the algebra $ sp(4)$ has rank two.

\abz Few comments are now in order. First of all, as follows from
\eqref{bhussym}, $\tau_\mu$ symmetry makes it possible to set one of the
$AdS_4$ invariants to  1, 0 or -1.
As we will seen it gives a black hole two kinematic parameters one
of which can be always taken discrete by diffeomorphism. Another
observation is that the Kerr--Schild vectors $l^{+-}_{\al\dal}$
and $l^{-+}_{\al\dal}$ may not exist for some values of $AdS_4$ invariants.
Indeed, consider the case with
$K_{A}{}^{C}K_{C}^{B}\sim \gd_{A}{}^{B}$ where
\be
C_4=C_2^2\,.
\ee
It is easy to see, that in this case
\be \label{mostsym}
\kappa_{\alpha\gamma}V^{\gamma}{}_{\dot\beta} =  \bar\kappa_{\dot\gamma\dot\beta}V_{\alpha}{}^{\dot\gamma}.
\ee
Direct consequence of \eqref{mostsym} is
\be
\kappa^2 = \bar\kappa^2, \qquad \G = \bar \G.
\ee
From \eqref{kap^2} and \eqref{mostsym} it follows that
\be \label{0kap^2}
K_{AC}K^C{}_B  = C_2 \left(
\begin{array}{cc}
\varepsilon_{\alpha\beta} &0 \\
0 & \varepsilon_{\dot\alpha\dot\beta}
\end{array}
\right)\,.
\ee
Using the definition \eqref{def-V+-} and \eqref{Pi-xieta} we have
\be\label{V-decomp}
V_{\al\dal}=V^{+}_{\al\dal}+V^{-}_{\al\dal}+V^{+-}_{\al\dal}+V^{-+}_{\al\dal}\,.
\ee
Substituting \eqref{V-decomp} into \eqref{mostsym} and using
\eqref{Vxieta}, \eqref{F-xieta} we find
\be \label{msym V}
V_{\alpha\dot\alpha}^{+-}  =  V_{\alpha\dot\alpha}^{-+} = 0.
\ee
Then from (\ref{ll-def}) it follows that
\be
l_{\alpha\dot\alpha}^{+-}\to\infty\q
l_{\alpha\dot\alpha}^{-+}\to\infty\,.
\ee
Moreover, taking  into account \eqref{vac-vpm} we obtain $I_2=0$. As
we will see later this case provides a black hole with vanishing
rotation parameter and only one non-zero invariant
$
C =   I_1\,,
$
while $I_2 = 0$.

\section{Black hole unfolded system}\label{FDA}

The equations \eqref{vac1-ads}--\eqref{vac2c-ads} admit a natural
deformation of the  $AdS_4$ unfolded system, that preserves its
Killing and Maxwell properties,  {\it i.e.}, we require the
deformed unfolded system to be built of  Killing vector and
source-free Maxwell tensor. As we will see this deformation
describes generic $AdS_4$ black hole.

\abz  Let us relax the equation (\ref{rho-ads}) for
$\rho$ in \eqref{vac1-ads} by allowing it to be an arbitrary function
of $\G$ and $\bar\G$
\be \label{rhoF}
\rho = \rho (\G,\bar\G)\,.
\ee
In this case the consistency condition for the system
\eqref{vac1-ads}--\eqref{vac2c-ads}  turns out to be very restrictive.
Solving Bianchi identities for \eqref{vac1-ads}--\eqref{vac2c-ads}
and taking \eqref{rhoF} into account we find the following most general
solution for $\rho$
\be \label{rho-sol}
\rho = \M - \lambda^2 \G^{-3} -\e\, \bar\G\,,
\ee
where $\M$ and $\e$ are, respectively, arbitrary  complex and real
parameters.

\abz As a result the complete consistent deformed unfolded
equations read
\begin{eqnarray}
\fD\fV_{\alpha\dot\alpha} & =& \frac12\rho\,
\fh^{\gamma}{}_{\dot\alpha}\F_{\gamma\alpha}+\frac12 \bar{\rho}\, \fh_{\al}{}^{\dot\gamma}\bar{\F}_{\dal\dgga}\,, \label{FDA1} \\
\fD\F_{\alpha\alpha} &=&-\frac{3}{2\G}
\fh^{\beta\dot\gamma}\fV^{\beta}{}_{\dot\gamma}\F_{(\beta\beta}\F_{\alpha\alpha)}\,, \label{FDA2} \\
\fD\bar{\F}_{\dot\alpha\dot\alpha} &=&-\frac{3}{2\bar{\G}}
\fh^{\gamma\dot\beta}\fV_{\gamma}{}^{\dot\beta}\bar{\F}_{(\dot\beta\dot\beta}\bar{\F}_{\dot\alpha\dot\alpha)}\,,\label{FDA2c}
\end{eqnarray}
with the following curvature two-forms
\begin{eqnarray}
\mathcal{R}_{\alpha\alpha} &=&
\f{\lambda^2}{2}\mathbf{H}_{\alpha\alpha}
-\frac{3(\M-\e\,\bar\G)}{4\G}\mathbf{H}^{\beta\beta}\F_{(\beta\beta}\F_{\alpha\alpha)}
+\frac{\e}{4}\, \bar{\mathbf{H}}^{\dot\beta\dot\beta}\bar\F_{\dot\beta\dot\beta}\F_{\alpha\alpha} \label{FDA3}\,,\\
\mathcal{\bar R}_{\dot\alpha\dot\alpha} &=&
\f{\lambda^2}{2}\mathbf{\bar H}_{\dot\alpha\dot\alpha}
-\frac{3(\overline\M-\e\,\G)}{4\bar\G}\bar{\mathbf{H}}^{\dot\beta\dot\beta}\bar\F_{(\dot\beta\dot\beta}\bar\F_{\dot\alpha\dot\alpha)}
+\frac{\e}{4}\,\mathbf{H}^{\beta\beta}\F_{\beta\beta}\bar\F_{\dot\alpha\dot\alpha}\label{FDA3c}\,,\\
\fD\fh_{\alpha\dot\alpha}&=&0\,, \label{FDAh}
\end{eqnarray}
and
\begin{eqnarray}
\rho &=& \M - \lambda^2 \G^{-3} -\e \,\bar\G, \quad
\bar\rho = \overline\M - \lambda^2 \bar\G^{-3} -\e\, \G \label{FDA4} \\
\quad \G &=& (-\F^2)^{1/4}, \quad \bar\G = (-\bar\F^2)^{1/4}, \label{FDA5}
\end{eqnarray}
where $\mathbf{H}^{\alpha\alpha}$ and $\mathbf{\bar
H}^{\dot\alpha\dot\alpha}$ are defined by \eqref{Hcap},
$\mathcal{R}_{\alpha\alpha}$ and $\mathcal{\bar
R}_{\dot\alpha\dot\alpha}$ are the curvatures \eqref{RHC} and
\eqref{RHCc}. For $\G$ and $\bar\G$ one finds the same consequence
as \eqref{dG}
\be \label{dG2}
d\G=-\frac12
\fh^{\alpha\dot\alpha}\fV^{\alpha}{}_{\dot\alpha}F_{\alpha\alpha},
\qquad d\bar\G=-\frac12
\fh^{\alpha\dot\alpha}\fV_{\alpha}{}^{\dot\alpha}\bar{F}_{\dot\alpha\dot\alpha}\,.
\ee

\abz The last term in \eqref{FDA3} and
\eqref{FDA3c} has the form of energy-momentum tensor for Maxwell
field and is invariant under $U(1)$ rotations \eqref{se-m}. Then
the integration constant $\e$ is interpreted as the sum of squares
of the electric and magnetic charges and can be written as $\e =
2(e^2+\g^2)$.

\abz We call the system \eqref{FDA1}--\eqref{FDA5} black hole
unfolded system (BHUS). (Note, that
the BHUS of \cite{DMV} is a particular case
of \eqref{FDA1}--\eqref{FDA5} with $\M=\overline{\M}$ and
$\e=0$.) The Weyl tensor it yields is of Petrov D-type.
Comparing \eqref{FDA3} and \eqref{FDA3c} with \eqref{RHC} and
\eqref{RHCc} we find out that the deformation \eqref{rhoF} of
$AdS_4$ algebra leads to vacuum Maxwell-Einstein equations, with
  the Maxwell tensor $\F_{\alpha\alpha}$ and
Weyl tensor given by
\be \label{weyl}
C_{\alpha\alpha\alpha\alpha}= -\frac{6(\M-\e\,\bar\G)}{\G}\F_{\alpha\alpha}\F_{\alpha\alpha}\,.
\ee
It follows from \eqref{bhussym} and \eqref{parity} that it is
$\tau_\mu-$invariant in accordance with \cite{Carter2} where  it
was shown that the Papapetrou's result concerning additional discrete
symmetry can be generalized to non-empty spaces with the matter
tensor invariant under simultaneous inversion of the time and
axial angle and that this holds automatically in the case of
source-free electromagnetic field.

\abz By analogy with the $AdS_4$ case, using
\eqref{FDA1}--\eqref{FDA2c} one can straightforwardly check that
the following expressions conserve  in BHUS
\begin{eqnarray}
\mathcal{I}_1 &=& \fV^2-\M\G - \overline\M\bar\G -\frac{\lambda^2}{2}\left(\frac{1}{\G^2}+\frac{1}{\bar\G^2}\right) +\e\,\G\bar\G\,, \label{FI1} \\
\mathcal{I}_2 &=&
\frac{1}{\G^3\bar{\G}^3}\fV^{\alpha\dot\alpha}\fV^{\alpha\dot\alpha}\F_{\alpha\alpha}
\bar\F_{\dot\alpha\dot\alpha} -
2\left(\frac{\M}{\G}+\frac{\overline\M}{\bar\G}\right) -
\mathcal{I}_1 \left(\frac{1}{\G^2}+\frac{1}{\bar\G^2}\right)
-\frac{\lambda^2}{4}\left(\frac{1}{\G^4}+\frac{1}{\bar\G^4}\right)
- \frac{3\lambda^2}{2\G^2\bar{\G}^2}.\qquad \label{FI2}
\end{eqnarray}
\[
d\mathcal{I}_1=d\mathcal{I}_2=0\,.
\]
In other words $\mathcal{I}_1$ and $\mathcal{I}_2$ are the first
integrals in the unfolded system.

\abz Remarkably, all the differential and algebraic properties of
BHUS literally coincide with those of the vacuum $AdS_4$ system of
Section \ref{AdS} with \eqref{vac-FI1}, \eqref{vac-FI2} replaced by \eqref{FI1},
\eqref{FI2} and \eqref{vac-Deltar}, \eqref{vac-Deltay} by
\begin{eqnarray}
\hat\Delta_r &=& 2\mathbf{M}r + r^2(\lambda^2r^2+\mathcal{I}_1) + \frac12(-\e+\frac{\mathcal{I}_2}{2}), \label{Deltar} \\
\hat\Delta_y &=& 2\mathbf{N}y + y^2(\lambda^2y^2-\mathcal{I}_1) +
\frac12(\e+\frac{\mathcal{I}_2}{2})\,, \label{Deltay}
\end{eqnarray}
where
\be
\mathbf{M} = \frac12(\M+\overline\M)\q \mathbf{N} =
\frac{1}{2i}(\M-\overline\M)\,.
\ee
In Subsection \ref{coord-carter} we will see that $\mathbf{M}$ and
$\mathbf{N}$ can be interpreted as mass and NUT-parameter of a
black hole. Analogously, one can build Kerr--Schild vectors in the
system using the same projector procedure. The differential
properties of these Kerr--Schild vectors are given by
\eqref{DeA}--\eqref{fegec} with the function $\rho$ of the form
\eqref{FDA4}. Using \eqref{FI1} and \eqref{FI2} and making the
change of variables \eqref{can-coord} we obtain
\begin{eqnarray}
(\fV^+\fV^-) &=& \frac{\hat\Delta_r}{r^2+y^2}, \label{vpp}\\
(\fV^{+-}\fV^{-+}) &=& -\frac{\hat\Delta_y}{r^2+y^2}, \label{vpm}
\end{eqnarray}

\abz Vector potentials for Maxwell tensor and its Hodge dual again
read as
\be \label{A1234}
A_{1,2}=\frac{r}{r^2+y^2}\hat\E_{1,2}, \quad A_{3,4}=\frac{y}{r^2+y^2}\hat\E_{3,4},
\ee
where
\be
\hat\E_I =\frac12  \hat e_{I,\alpha\dot\alpha}
\fh^{\alpha\dot\alpha}=(\hat{K}, \hat{N}, \hat{L}^{+-}, \hat{L}^{-+})\,. \label{eform-def}
\ee
Here
\be\label{KS-vectors}
\hat e_{I,\alpha\dot\alpha}=(\hat{k}_{\al\dal}, \hat{n}_{\al\dal},
\hat{l}^{+-}_{\al\dal}, \hat{l}^{-+}_{\al\dal})
\ee
are defined the same way as in \eqref{eA-def}. Recall, that
(un)hatted quantities are associated with  the (un)deformed
unfolded system.

\abz The following relations remain true
\be \label{e1-e2}
\hat{K}-\hat{N} = \frac{2(r^2+y^2)}{\hat\Delta_r}dr, \qquad \hat{L}^{+-}-\hat{L}^{-+} =
\frac{2(r^2+y^2)}{i\hat\Delta_y}dy\,.
\ee
Again, $\hat{K}$ and $\hat N$ are real one-forms, while $\hat
L^{+-}$ and $\hat L^{-+}$ are complex conjugated.
Finally, $A_{1,2}$ generate the same Maxwell tensor $F_{ij}$ and
$A_{3,4}$ generate $*F_{ij}$.

\abz The free and deformed unfolded systems are similar in many
respects. In particular, they have the same number of first
integrals, Kerr-- Schild vectors, both have source-free Maxwell
and closed Killing--Yano tensors. At $\mathcal{M}=0$, $\e=0$ the
two systems just coincide. All this suggests that there should be
some integrating flow with respect to the parameters $\M$ and $\e$
that maps  one system to another. The existence of such an
integrating flow is also natural in the context of expected
relationship of the proposed construction with yet unknown black
hole solution of the nonlinear higher spin gauge theory. Indeed,
the integrating flow approach we are about to explore is to large
extend analogous to the integrating flow in higher spin gauge
theory \cite{Gol} that maps solutions on  nonlinear higher spin
equations to those of free higher spin equations.

\section{Integrating flow} \label{IF}

Now we are in a position to construct  the integrating flows with
respect to $\M$ and $\e$. The  benefit  of using the integrating
flows which are first order differential equations with respect to
the modules of the black hole solution is that by solving these
equations it is easy to reconstruct the black hole solutions in
terms of the initial data that describe the vacuum $AdS_4$
geometry. In other words, the idea is to obtain complicated black
hole solutions of Einstein theory as solutions of simple and
easily integrable flow equations whose form is fixed by the formal
consistency conditions
\be\label{consist}
\left[ \frac{\p}{\p \M}, \frac{\p}{\p x^m} \right] =  \left[
\frac{\p}{\p \overline\M}, \frac{\p}{\p x^m} \right] = \left[
\frac{\p}{\p \e}, \frac{\p}{\p x^m} \right] = \left[ \frac{\p}{\p
\M}, \frac{\p}{\p \e} \right]= \left[ \frac{\p}{\p \M},
\frac{\p}{\p \overline\M} \right]=0\
\ee
with equations \eqref{FDA1}--\eqref{FDA2c}, \eqref{FDAh}.

\abz We require the Maxwell tensor  to be constant along the
flows and hence so are $\G$ and $\bar\G$ (equivalently, $r$ and
$y$)
\be \label{dM0}
\frac{\p}{\p \M} \F_{\alpha\alpha} = \frac{\p}{\p \overline\M}
\F_{\alpha\alpha} = \frac{\p}{\p \e} \F_{\alpha\alpha} = 0\,.
\ee
Although the requirement \eqref{dM0} is not necessary,  it
drastically simplifies the analysis.  Also it is natural in a
sense that known examples of black holes curvature tensors can be
reduced to the form that agrees with \eqref{dM0}. Indeed, the
condition \eqref{dM0} turns out to be consistent with
\eqref{FDA1}--\eqref{FDA2c}. Note, that the integrating flow with
respect to $\overline\M$ can be obtained by complex conjugation of
the $\M$-flow\footnote{Strictly speaking it is true given the
reality conditions \eqref{real-cond} imposed.}. Leaving the detail
of derivation for Appendix \ref{App-C}, the final result for the
integrating flows is
\be\label{M-flow}
\p_{\M}\fV_{\al\dal}=\sum_{I=1}^{4}\phi_I\hat{e}_{I, \al\dal}\q
\p_{\M}\fh_{\al\dal}=\sum_{I=1}^{4}\phi_I\hat{e}_{I,
\al\dal}\hat\E_I
\ee
and
\be\label{q-flow}
\p_\e\fV_{\al\dal}=\sum_{I=1}^{4}\psi_I\hat{e}_{I, \al\dal}\q
\p_{\e}\fh_{\al\dal}=\sum_{I=1}^{4}\psi_I\hat{e}_{I,
\al\dal}\hat\E_I\,,
\ee

\abz For the generic case with arbitrary complex $\M$ the
functions $\phi_I$ and $\psi_I$ read
\be\label{1}
\phi_1=\f{\G+\bar\G}{4}\al_1(r)\q
\phi_2=\f{\G+\bar\G}{4}\al_2(r)\,,
\ee
\be\label{2}
\phi_3=\f{\G-\bar\G}{4}\gb_1(y)\q \phi_4=\f{\G-\bar\G}{4}\gb_2(y)
\ee
and
\be\label{3}
\psi_1=-\f{\G\bar\G}{4}\theta_1(r)\q
\psi_2=-\f{\G\bar\G}{4}\theta_2(r)\,,
\ee
\be\label{4}
\psi_3=-\f{\G\bar\G}{4}\vartheta_1(y)\q
\psi_4=-\f{\G\bar\G}{4}\vartheta_2(y)\,,
\ee
where functions $\al$, $\gb$, $\theta$, $\vartheta$ satisfy the constraints
\be\label{cnstr-1}
\al_1(r)+\al_2(r)=\gb_1(y)+\gb_2(y)=1\,,
\ee
\be\label{cnstr-2}
\theta_1(r)+\theta_2(r)=1-\f{\gga}{2}\q
\vartheta_1(y)+\vartheta_2(y)=1+\f{\gga}{2}
\ee
and
\be\label{cnstr-3}
\al_1(r)\theta_2(r)=\al_2(r)\theta_{1}(r)\q
\gb_1(y)\vartheta_2(y)=\gb_2(y)\vartheta_1(y)\,,
\ee
where $\gga$ is an arbitrary real constant. In addition, we assume that
$\al$, $\gb$, $\theta$, $\vartheta$, $\gga$ are $\M$- and
$\e$-independent. The reality condition for the gravitational
fields requires in addition
\be\label{real-cond}
\gb_1(y)=\gb_2(y)=\f12\q
\vartheta_1(y)=\vartheta_2(y)=\f12+\f{\gga}{4}\,.
\ee
Note that sometimes it is
useful  to deal with complex metric (particularly, with
its double Kerr--Schild form).

\abz The flows \eqref{M-flow}, \eqref{q-flow} also imply
\be
\p_\M\mathcal{I}_1=\p_{\M}\mathcal{I}_2=0\q \p_{\e}\mathcal{I}_1=0
\q \p_\e\mathcal{I}_2=\gga\,.
\ee

\abz Let us emphasize, that the arbitrary functions
$\al_{1,2}(r)$, $\gb_{1,2}(y)$ and $\gga$ represent the pure gauge
ambiguity. These arise as integration constants in the integrating flow
equations (see Appendix \ref{App-B}) and restricted only by the
reality condition \eqref{real-cond} if necessary. This gauge
ambiguity encodes in a rather nontrivial way the ambiguity in the
choice of one or another coordinate system, giving the
integrating flow approach  the wide area of applicability.

\abz Note, that the case of $\M=\overline\M$ and $\e=0$ was
considered in \cite{DMV} within the same unfolded approach.
However, the simple Kerr--Schild shift used there to map BHUS into
the free $AdS_4$ system does not work when $\M\neq\pm\overline\M$.
The reason is that the Kerr--Schild shift used in \cite{DMV} is
not compatible with the reality of metric in this case. The answer
in terms of the integrating flow reduces to different constraint
conditions for flow functions (see Appendix \ref{App-B}).

\abz In the case $\M=\overline\M$, we perform different flow
integration, than that of general complex $\M$, resulting in
notable simplification of \eqref{M-flow}, \eqref{q-flow}. In this
case we fix the gauge freedom as follows
\be\label{gauge-1}
\phi_1=\f12(\G+\bar\G)\q \phi_2=\phi_3=\phi_4=0\,,
\ee
\be\label{gauge-2}
\psi_1=-\f12\G\bar\G\q \psi_2=\psi_3=\psi_4=0\,.
\ee
This in turn implies
\be
\p_\M\mathcal{I}_1=\p_{\M}\mathcal{I}_2=0\q \p_{\e}\mathcal{I}_1=0
\q \p_\e\mathcal{I}_2=-2\,.
\ee
Note, that the difference between the cases with
$\M=\pm\overline\M$ and $\M\neq\pm\overline\M$ arises as a
consequence of the possibility to replace the two flows with
respect to $\M$ and $\overline\M$ in the latter case by a single
flow with respect to $\M$ in the former. (For more detail see
Appendix \ref{App-B}.)

\abz Let us stress, that the condition \eqref{dM0} has been
extensively used in the derivation of \eqref{M-flow},
\eqref{q-flow}. The obtained integrating flows  can be explicitly
integrated, giving the $AdS_4$ covariant and coordinate-free
description of a black hole metric as we demonstrate in the next
section.

\abz The following comment is now in order. In the derivation of
the integrating flow equations we have fixed the gauge freedom. In
principle, this could have been done in variety of ways keeping
one or another amount of arbitrary gauge parameters. Our strategy
was to leave those as few as possible, though still enough to
encompass the most important representations, such as
Kerr--Schild, double Kerr--Schild and the generalized
Carter--Plebanski. In principle one can think of further
generalization of the form of the integrating flows to describe
even more general forms of the black hole solutions.

\section{Flow integration with $AdS_4$ initial data}\label{S-BHUS}

\subsection{Solution for Kerr--Schild vectors} \label{sec:null-expr}

To restore BHUS frame fields from integrating flow equations it is
convenient to start with Kerr--Schild vectors. Let us start with
the generic case of arbitrary complex $\M$.

\abz It is convenient to restrict the integrating flow gauge
parameters \eqref{cnstr-1}--\eqref{cnstr-3} as follows
\be\label{func-fix}
\al_1(r)=\theta_1(r)\,,\quad\al_2(r)=\theta_2(r)\,,\quad
\gb_1(y)=\vartheta_1(y)\,,\quad
\gb_2(y)=\vartheta_2(y)\,,\quad\gga=0\,.
\ee
Note that this gauge choice is compatible with the reality
condition \eqref{real-cond} only if $\gb_1=\gb_2=1/2$.
As already mentioned, the other choices may still
be useful to incorporate the double Kerr--Schild form of the
complex black hole metric.

\abz As explained in Appendix \ref{App-C},
 the flow equations \eqref{M-flow}, \eqref{q-flow} for the Kerr--Schild vectors
give
\begin{eqnarray}
\hat k_{\alpha\dot\alpha} &=& k_{1,\alpha\dot\alpha} \left(
\frac{\Delta_r}{\hat\Delta_r}\right)^{\alpha_2},
\hat n_{\alpha\dot\alpha} = n_{\alpha\dot\alpha} \left( \frac{\Delta_r}{\hat\Delta_r}\right)^{\al_1},  \label{e-vec1} \\
\hat l^{+-}_{\alpha\dot\alpha} &=& l^{+-}_{\alpha\dot\alpha}
\left( \frac{\Delta_y}{\hat\Delta_y}\right)^{\gb_2}, \quad \hat
l^{-+}_{\alpha\dot\alpha} = l^{-+}_{\alpha\dot\alpha} \left(
\frac{\Delta_y}{\hat\Delta_y}\right)^{\gb_1}, \label{e-vec2}
\end{eqnarray}
where the unhatted quantities  correspond to the initial data of the
$AdS_4$ unfolded system
 \eqref{vac1-ads}--\eqref{vac2c-ads}
\be
\Delta_{r,y} =  \left.\hat\Delta_{r,y}\right|_{\M,\overline\M,\e
=0}\q e_{I ,\al\dal} =  \left.\hat{e}_{I
,\al\dal}\right|_{\M,\overline\M,\e =0}\,.
\ee
Analogously,  the integration of the Kerr--Schild one-forms
$\hat\E_I$ \eqref{eform-def} gives
\be
\hat K = K -\alpha_2(r) \frac{\hat\Delta_r-\Delta_r}{\hat\Delta_r}
(K-N)\,,\quad \hat N= N -\alpha_1(r)
\frac{\hat\Delta_r-\Delta_r}{\hat\Delta_r} (N-K)\,,
\label{e-form1}
\ee
\be
\hat L^{+-} = L^{+-} -\gb_2(y)
\frac{\hat\Delta_y-\Delta_y}{\hat\Delta_y} (L^{+-}-L^{-+}), \quad
\hat L^{-+}= L^{-+} -\gb_1(y)
\frac{\hat\Delta_y-\Delta_y}{\hat\Delta_y} (L^{-+}-L^{+-})\,.
\label{e-form2}
\ee
Note, that
\begin{eqnarray}
\alpha_1K + \alpha_2N &=& \alpha_1\hat K + \alpha_2\hat N, \\
\gb_1 L^{+-} + \gb_2 L^{-+} &=& \gb_1\hat L^{+-} + \gb_2\hat
L^{-+}.
\end{eqnarray}
Note also that $\hat L^{+-}$ and $\hat L^{-+}$ are complex
conjugated only if $\gb_1=\gb_2=\f12$. The first integrals for
that particular gauge choice \eqref{func-fix} coincide with those
of $AdS_4$
\be\label{int-equal}
\mathcal{I}_1=I_1\q \mathcal{I}_2=I_2\,.
\ee

\abz Now consider the special case of  $\M=\overline\M$. Using the
simplest gauge choice for the flow coefficients \eqref{gauge-1},
\eqref{gauge-2}  the integration of the vierbein field gives  the
Kerr--Schild form
\be
\hat k_{\al\dal}=k_{\al\dal}\q \hat K=K\,,
\ee
\be\label{KS-tetrad}
\fh_{\al\dal}=h_{\al\dal}+\f12\Big(M(\G+\bar\G)-\e\G\bar\G\Big)k_{\al\dal}K
\ee
with
\be
\mathcal{I}_1=I_1\q \mathcal{I}_2=I_2-2\e\,.
\ee
Let us stress again that for $C_4=C_2^2$ the Kerr--Schild vectors
$l^{+-}_{\al\dal}$ and $l^{-+}_{\al\dal}$ are ill-defined already
in the vacuum $AdS_4$ geometry and hence the vectors  $\hat
l^{+-}_{\al\dal}$ and $\hat l^{-+}_{\al\dal}$ can not be expressed
in terms of their $AdS_4$ counterparts.

\abz The following comment is now in order. The $AdS_4$ unfolded
system \eqref{vac1-ads}--\eqref{vac2c-ads} admits well defined
flat limit $\gl\to 0$. Therefore one can integrate BHUS with
Minkowski space-time set of initial data by introducing an
additional flow with respect to cosmological constant $\gl^2$. In
this case BHUS first integrals would be related to invariants of
Poincar\'e algebra
 and the solution
would be written in Minkowski covariant way. We, however, prefer
to work in $AdS_4$ covariant way rather than in Poincar\'e.

\subsection{$AdS_4$ covariant form of a black hole metric}\label{BH-metric}

The obtained  Kerr--Schild vectors and one-forms \eqref{e-vec1},
\eqref{e-vec2}, \eqref{e-form1}, \eqref{e-form2} allow us to
reconstruct black hole vierbein and metric. Let us consider the
generic case of arbitrary complex $\M$. To reproduce the metric we
use the following identity
\be \label{int vierbein}
\fh_{\alpha\dot\alpha} = \frac{2}{(\hat k \hat n)}(\hat
k_{\alpha\dot\alpha}\hat N + \hat n_{\alpha\dot\alpha}\hat K) +
\frac{2}{(\hat l^{+-} \hat l^{-+})}(\hat
l^{+-}_{\alpha\dot\alpha}\hat L^{-+} + \hat
l^{-+}_{\alpha\dot\alpha}\hat L^{+-})\,,
\ee
which arises as a consequence of the completeness relation for
two-component spinors
\be
\gep_{\al\gb}\gep_{\dal\dgb}=\f{1}{(\hat k\hat n)}(\hat
k_{\al\dal}\hat n_{\gb\dgb}+\hat n_{\al\dal}\hat
k_{\gb\dgb})+\f{1}{(\hat l^{+-}\hat l^{-+})}(\hat
l^{+-}_{\al\dal}\hat l^{-+}_{\gb\dgb}+\hat l^{-+}_{\al\dal}\hat
l^{+-}_{\gb\dgb})\,,
\ee
which can be straightforwardly checked using \eqref{Vxieta}.
Substituting \eqref{e-vec1}, \eqref{e-vec2}, \eqref{e-form1},
\eqref{e-form2} into \eqref{int vierbein} we restore vierbein  in
a coordinate-independent way.

\abz The metric is
\be
ds^2 = \frac12 \fh_{i\alpha\dot\alpha}\fh_{j}{}^{\alpha\dot\alpha}dx^idx^j.
\ee
Substituting \eqref{int vierbein} we obtain coordinate-independent
representation for the metric in the form
\be \label{int metric}
ds^2 =\frac{\hat\Delta_r}{r^2+y^2}\hat K\hat N -
\frac{\hat\Delta_y}{r^2+y^2}\hat L^{+-}\hat L^{-+},
\ee
where $\hat\Delta_r$ and $\hat\Delta_y$ are defined in
\eqref{Deltar} and \eqref{Deltay}, respectively. Now with the help
of \eqref{e-form1} and \eqref{e-form2} one can rewrite the metric
in terms of $AdS_4$ fields \eqref{vac1-ads}--\eqref{vac2c-ads}
\begin{align} \label{sol-cbh}
ds^2 &= ds_0^2 + \frac{2\bM r-\e/2}{r^2+y^2} (\alpha_1(r) K+
\alpha_2(r) N)^2 -
\frac{2\bN y+\e/2}{r^2+y^2} (\gb_1(y) L^{+-}+ \gb_2(y) L^{-+})^2 \notag \\
&+
4\alpha_1(r)\alpha_2(r)\frac{r^2+y^2}{\Delta_r\hat\Delta_r}(2\bM
r-\e/2)dr^2 -
4\gb_1(y)\gb_2(y)\frac{r^2+y^2}{\Delta_y\hat\Delta_y}(2\bN
y+\e/2)dy^2,
\end{align}
where $\alpha_1(r)+\alpha_2(r)=\gb_1(y)+\gb_2(y)=1$ and $ds^2_0$
is the background $AdS_4$ metric which can be represented in the
form analogous to (\ref{int metric})
\be \label{bgmetric}
ds_0^2 = \frac{\Delta_r}{r^2+y^2}KN -
\frac{\Delta_y}{r^2+y^2}L^{+-}L^{-+}.
\ee
The reality condition for \eqref{sol-cbh} requires
\be\label{real}
\gb_1=\gb_2=\f12\,.
\ee
The case of complex metric with, say, $\al_1=\gb_1=1$,
$\al_2=\gb_2=0$ yields the double Kerr--Schild form of \eqref{sol-cbh}
\be \label{a1a3metric}
ds^2 = ds_0^2 +  \frac{2r}{r^2+y^2}\left(\mathbf{M} -
\frac{e^2+\g^2}{2r}\right) KK -
 \frac{2y}{r^2+y^2}\left(\mathbf{N} + \frac{e^2+\g^2}{2y}\right)
 L^{+-}L^{+-}\,,
\ee
which might be useful as it satisfies both linearized and nonlinear
Einstein--Maxwell equations\footnote{Note, that \eqref{a1a3metric}
becomes  real upon Wick rotation to (2,2) signature.}.

\abz To reveal physical meaning of the metric \eqref{sol-cbh} let
us first recall its parameter space. There are three parameters
$\mathbf{M}$, $\mathbf{N}$,  $\e$ and $\gl$ that cannot be
redefined by diffeomorphisms as they enter the Riemann curvature
tensor in \eqref{FDA3}, \eqref{FDA3c}. Then, there are two
parameters associated with the first integrals \eqref{int-equal}
and expressed via $AdS_4$  invariants by \eqref{ads-c1},
\eqref{ads-c2}. These are so called kinematical parameters one of
which can be always chosen to be -1, 0 or 1. Indeed, the obtained
integrating flow transforms BHUS into $AdS_4$ global symmetry
parameter equation \eqref{ads const1}, \eqref{ads const2} which is
invariant under rescaling  \eqref{scaling}. This allows us to set,
say, $C_2=\pm 1$ or 0.

\abz As a result, the diffeomophism invariant black hole parameter
space consists of three curvature parameters $\mathbf{M}$,
$\mathbf{N}$, $\e$ (and $\gl$) and two kinematical ones,  discrete
$C_2$ and continuous  $C_4$.

\abz Let us show that \eqref{sol-cbh} is nothing but a coordinate-independent
realization of the  $AdS_4$-Kerr--Newman-Taub-NUT
solution originally discovered by Carter \cite{Carter4} and
Plebanski \cite{Pleb}. To this end we choose  certain
two-parametric coordinate realization of $AdS_4$ space that covers
whole range of  values of invariants for a particular $AdS_4$
Killing vector and calculate the resulting metric \eqref{sol-cbh}.

\section{Coordinate realization of $AdS_4$ background} \label{coord-AdS}

Following \cite{Carter4} it is convenient to specify $AdS_4$
metric in certain two-parametric form. Using the coordinate system
$\{\tau, \psi, r, y\}$ $AdS_4$ metric can be written down in the
form
\be \label{bgmetric-coord}
ds^2_0 = \frac{\Delta_r}{r^2+y^2} (d\tau +y^2d\psi)^2 -
\frac{\Delta_y}{r^2+y^2} (d\tau -r^2d\psi)^2 -
\frac{r^2+y^2}{\Delta_r} dr^2 - \frac{r^2+y^2}{\Delta_y} dy^2\,,
\ee
where
\be  \label{Delta-bg}
\Delta_r = r^2(\lambda^2r^2 + \epsilon) + a^2, \quad \Delta_y =
y^2(\lambda^2y^2 - \epsilon) +  a^2.
\ee
The parameters $a$ and $\gep$ that enter the $AdS_4$ metric
\eqref{bgmetric-coord} as pure gauge arbitrary
constants\footnote{$a^2$ can be chosen to be negative as well.}
become the Carter--Plebanski black hole kinematic parameters upon
$AdS_4$ deformation. The metric verifies $AdS_4$ Einstein
equations and provides
\be
R_{ij} = 3\lambda^2 g_{ij}\,.
\ee
Now, the $\f{\p}{\p t}$ Killing vector
\be\label{Killing}
V^i = \{1, 0, 0, 0\}
\ee
renders via \eqref{vac1-ads} the source-free Maxwell two-form
\be\label{Maxwell}
F=\frac{1}{(r^2+y^2)^2}( (d\tau+y^2\psi) \wedge (r^2-y^2)dr +
2(d\tau-r^2\psi)\wedge rydy)\,,
\ee
that can be generated by the vector potential one-form
\be
F=\text{d}A\q A=\frac{r}{r^2+y^2}(d\tau +y^2\psi).
\ee
Maxwell tensor \eqref{Maxwell} along with the Killing vector
\eqref{Killing} fulfill  $AdS_4$ unfolded equations
\eqref{vac1-ads}--\eqref{vac2c-ads}. Coordinates $r$ and $y$ can be
checked to coincide with the canonical coordinates
\eqref{can-coord}.

\abz Using \eqref{kn-def} and \eqref{ll-def} we find Kerr--Schild
one-forms in the specified coordinates
\begin{eqnarray} \label{bg-e-set}
K &=& d\tau + y^2 d \psi + \frac{r^2+y^2}{\Delta_r}dr, \quad
N = d\tau + y^2 d \psi  - \frac{r^2+y^2}{\Delta_r}dr, \label{E1-bg} \\
L^{+-} &=& d\tau - r^2 d \psi + \frac{r^2+y^2}{i\Delta_y}dy, \quad
L^{-+} = d\tau - r^2 d \psi - \frac{r^2+y^2}{i\Delta_y}dy\,.
\label{E3-bg}
\end{eqnarray}
Note, that having expressions for background one-forms
\eqref{E1-bg} and \eqref{E3-bg} it is  easy  to calculate
coordinate form of Killing--Yano and closed Killing--Yano tensors \eqref{yka} and
\eqref{cky}, respectively
\begin{eqnarray}
Y &=& ydr\wedge(dt+y^2d\psi)  + rdy\wedge(dt-r^2d\psi) , \\
*\!Y &=& rdr\wedge(dt+y^2d\psi)  - ydy\wedge(dt-r^2d\psi) .
\end{eqnarray}

\abz The first integrals for the $AdS_4$ unfolded system
\eqref{vac-FI1}, \eqref{vac-FI2} amount to
\be\label{Int-a,e}
I_1=\gep\q I_2=4a^2
\ee
The obtained $AdS_4$ formulae lead to the Carter--Plebanski metric.

\section{Particular solutions}

\subsection{Carter--Plebanski solution} \label{coord-carter}

Consider the real case \eqref{real} of \eqref{sol-cbh} and let us
fix the gauge functions as
\be
\al_1(r)=\al_2(r)=\f12
\ee
Substituting \eqref{bgmetric-coord} and \eqref{E1-bg},
\eqref{E3-bg} into \eqref{sol-cbh} we obtain
\be \label{sol-rbh}
ds^2 = \frac{\hat\Delta_r}{r^2+y^2} (d\tau +y^2d\psi)^2 -
\frac{\hat\Delta_y}{r^2+y^2} (d\tau -r^2d\psi)^2 -
\frac{r^2+y^2}{\hat\Delta_r} dr^2 - \frac{r^2+y^2}{\hat\Delta_y}
dy^2\,,
\ee
where
\begin{eqnarray}
\hat\Delta_r &=& 2\mathbf{M}r- e^2-g^2 + r^2(\lambda^2r^2+\epsilon) + a^2, \label{sDeltar} \\
\hat\Delta_y &=& 2\mathbf{N}y +e^2+g^2  +
y^2(\lambda^2y^2-\epsilon) + a^2. \label{sDeltay}
\end{eqnarray}
(Recall, that $\e = 2(e^2+\g^2)$.) The metric \eqref{sol-rbh} is the
well-known Carter--Plebanski solution of vacuum Einstein--Maxwell
equations that describes Petrov D-type metric characterized by the
mass $\bM$, NUT-parameter $\bN$, electric and magnetic charges
$e$ and $\g$. It possesses two first integrals \eqref{FI1}, \eqref{FI2}
$a^2$ and $\gep$ one of which  can be chosen to be 1, 0 or -1. On account of
\eqref{int-equal} and \eqref{ads-c1}, \eqref{ads-c2} these
kinematical parameters are related to the $AdS_4$  invariants
\be
\gep=C_2\q 4\gl^2a^2=C_4-C_2^2\,.
\ee

\abz Since the coordinate realization of the $AdS_4$ metric
\eqref{bgmetric-coord} and  Killing vector \eqref{Killing} covers
the whole range of $AdS_4$ invariants \eqref{Int-a,e} it is shown
that BHUS with generic parameters describes Carter--Plebanski
solution \eqref{sol-rbh}. The parametric form \eqref{sol-cbh}
allows one to choose its different representations. In particular,
the double Kerr--Schild form \eqref{a1a3metric} is constructed
from  mutually orthogonal background null-vectors $k_{\al\dal}$,
$l^{+-}_{\al\dal}$ and depends on the deformation parameters
linearly. By an appropriate change of variables that changes the
metric signature one can obtain usual real form of the metric in
Minkowski signature \cite{Gibon2}.

\subsection{Kerr--Newman solution}

For physical applications it is instructive to consider the case
with vanishing NUT parameter $\bN=0$ that corresponds to real
$\M$. This case was considered in \cite{DMV} for $\e=0$ (Kerr
metric) within the unfolded dynamics approach. The case of $\e\neq
0$ corresponds to Kerr--Newman solution that describes  rotating
and electro-magnetically charged black hole (provided its charge
and rotation parameter are such, that the metric is free of naked
singularities). To describe this case one can simply set $\bN=0$
in \eqref{sol-cbh}. However as we have already seen, the
integrating flow with $\M=\overline\M$ admits a different
integration \eqref{KS-tetrad} giving rise to the simpler
Kerr--Schild form of a metric
\be
ds^2 = ds_0^2 + \frac{2\bM r-\e}{r^2+y^2} KK\,.
\ee
To associate $AdS_4$  invariants with the black hole rotational
parameter one can use the coordinate realization of \cite{Gib} for
$AdS_4$ metric and certain $AdS_4$ Killing vector (see \cite{DMV}
for detail) to find
\be
C_2=1+\gl^2 a^2\q C_4=C^2_2+4\gl^2 a^2\,.
\ee
The important particular case of static solution (charged
Schwarzschild black hole) arises if $a=0$, or equivalently
$
C_4=C^2_2\neq 0\,.
$

\section{Summary and discussion} \label{sum}
To conclude let us summarize the results obtained in this paper.
It is shown that a wide class of black hole metrics
(Carter--Plebanski) admits simple unfolded description in terms of
Killing and source-free Maxwell fields. The system is obtained as
a parametric deformation of $AdS_4$ global symmetry equation. Two
deformation parameters $\M\in \mathbb{C}$ and $\e\in \mathbb{R}$
are associated with black hole mass $\bM$=Re $\M$, NUT charge
$\bN$=Im $\M$ and electro-magnetic charges $\e=2(e^2+\g^2)$. Black
hole kinematic characteristics related to the angular momentum $a$
and the discrete parameter $\gep$ are expressed via two first
integrals of the unfolded system.

\abz It is shown that $AdS_4$ global symmetry equation and BHUS
are related by the integrating flows, that describe an evolution
with respect to black hole charges. The integrating flows allowed
us to describe the black hole vierbein, metric, Killing fields,
and other characteristics in the $AdS_4$ covariant and
coordinate-independent form. This was done by the straightforward
integration of the first order flow evolution equations with the
initial data corresponding to the $AdS_4$ vacuum space. One of the
consequences of this procedure is that black hole kinematic
parameters acquired invariant interpretation in terms of two
$AdS_4$  invariants. Let us stress, that the flow correspondence
between vacuum and black hole systems is similar to the one
studied in the nonlinear higher spin gauge theory \cite{PV,Gol}.

\abz We believe that the obtained results can have various
 useful applications. One of the most striking features
of the obtained description is that it does not refer to any
particular coordinate system. Many important algebraic objects
resided in black holes such as Killing--Yano tensors, Kerr--Schild
vectors acquire simple and natural origin in the proposed
formulation.

\abz One of the most straightforward applications could be the study
of fluctuations of different types of fields in the black hole geometry
in the unfolded dynamics approach.
Fortunately, the unfolded formulation of free fields of various types is
available in the literature (see e.g. \cite{Ann} for massless fields in $AdS_4$,
\cite{SV} for the case of scalar field of any mass in arbitrary dimension and
\cite{Gol,solv} for more references).

\abz Another intriguing development is to study a possible generalization
of the obtained results to the full nonlinear
higher spin gauge  theory that rests on the unfolded
formulation. Hopefully, the obtained results will allow us to challenge
this problem at least perturbatively.

\abz An interesting direction for the future study is to explore
the higher-dimensional generalization. It is well-known that black
holes in higher dimensions have  reacher properties than their
four-dimensional counterparts. In particular, in $d>4$ there are
black holes with non-spherical (ring) horizon topology
\cite{Reall}. Besides, the curvature tensor of higher-dimensional
black holes is not necessarily of generalized D-type
\cite{Pravdova} (e.g., in the case of black rings). Even though
the approach demonstrated in this paper is essentially
four-dimensional, we hope it can be extended to higher dimensions.
Let us note in this respect that the analysis of \cite{Frol},
where the hidden symmetries of higher-dimensional black holes were
discovered, suggests that the unfolded description of these black
holes is likely to be based on the differential forms of higher
ranks.

\abz An alternative possibility for a higher-dimensional
generalization of the obtained unfolded system is the analysis of
the black hole-like solutions in $Sp(2M)$ spaces with matrix
coordinates \cite{PRD, Vas-Sp, Plyush, DV}. Since these models
provide a higher-dimensional generalization of the spinor approach
used in this paper, one can speculate that such an extension may
be even simpler that in the usual tensorial setup.

\abz The last but not the least is to understand better the origin
of the integrating flow, which very likely is a manifestation of
some hidden higher dimensional and/or integrable structure in the
system.

\section*{Acknowledgement}
 This research was supported in part by INTAS Grant No
05-1000008-7928, RFBR Grant No 08-02-00963, LSS No 1615.2008.2
and  Alexander von Humboldt Foundation Grant PHYS0167.
A.M. acknowledges financial support from Dynasty Foundation
and Landau Scholarship Foundation.

\renewcommand{\theequation}{A.\arabic{equation}}
\renewcommand{\thesubsection}{A}
\renewcommand{\thesection}{Appendix}
\makeatletter \@addtoreset{equation}{subsection} \makeatother

\section{}

\subsection{Notations} \label{App-A}

Capital Latin letters $A,B,\dots$ label $Sp(4)$ vector ({\it
i.e.}, $4d$ Majorana spinor) indices, $A,B=1,\dots 4$. Indices
$i,j,\dots= 1,\dots 4$ are world (base), while $a,b,\dots =
1,\dots 4$ are fiber ones. Capital Latin indices from the middle
of alphabet $I,J = 1, \ldots,4$ are used to enumerate basis
null-vectors in integrating flow decomposition. To distinguish
between $AdS_4$ background and black hole quantities the latter
are endowed with hats.

\abz The analysis in four dimensions considerably simplifies in
spinor notation. Vector notation is translated  to the spinor one
and vice versa with the help of $\sigma$-matrices
($\sigma^0_{\alpha\dot\alpha}$ is the unit matrix and
 $\sigma^{1,2,3}_{\alpha\dot\alpha}$ are Pauli matrices) that obey the condition
\be
\eta_{ab} \,\sigma^a_{\alpha\dot\alpha}\sigma^b_{\beta\dot\beta} =
2\varepsilon_{\alpha\beta}\,\varepsilon_{\dot\alpha\dot\beta}\,,
\ee
where $\alpha,$ and $\dot\alpha=1,2$ are mapped to each other by
complex conjugation $\al\leftrightarrow\dal$. For a Lorentz vector
$V_a$ we have
\be
V_{\alpha\dot{\alpha}}=(\sigma^a)_{\alpha\dot{\alpha}}V_a\,,\quad
V_a =
\frac12(\sigma_a)^{\alpha\dot{\alpha}}V_{\alpha\dot{\alpha}}\,.
\ee
Spinor indices are raised and lowered  by the $sp(2)$
antisymmetric tensors $\varepsilon_{\alpha\beta}$ and
$\varepsilon_{\dot\alpha\dot\beta}$
\be
\xi_{\alpha}=\xi^{\beta}\varepsilon_{\beta\alpha}\,,\quad
\xi^{\alpha}=\varepsilon^{\alpha\beta}\xi_{\beta}\q
\bar\xi_{\dal}=\bar\xi^{\dgb}\varepsilon_{\dgb\dal}\,,\quad
\bar\xi^{\dal}=\varepsilon^{\dal\dgb}\bar\xi_{\dgb}\,,
\ee
where $\varepsilon_{12}=\varepsilon^{12}=1$,
$\varepsilon_{\alpha\beta}=-\varepsilon_{\beta\alpha}$,
$\varepsilon^{\alpha\beta}=-\varepsilon^{\beta\alpha}$.

\abz Lorentz irreducible spinor decompositions of the Maxwell and
Weyl tensors $F_{ab}$ and $C_{abcd}$ read, respectively, as
\be \label{sp2decomp}
F_{\alpha\dot{\alpha}\beta\dot{\beta}} =\varepsilon_{\alpha\beta}
\bar{F}_{\dot{\alpha}\dot{\beta}}+\varepsilon_{\dot{\alpha}\dot{\beta}}
F_{\alpha\beta}\q
C_{\alpha\dot{\alpha}\beta\dot{\beta}\gamma\dot{\gamma}\delta\dot{\delta}}
= \varepsilon_{\alpha\beta}\varepsilon_{\gamma\delta}
\bar{C}_{\dot{\alpha}\dot{\beta}\dot{\gamma}\dot{\delta}}
+\varepsilon_{\dot{\alpha}\dot{\beta}}\varepsilon_{\dot{\gamma}\dot{\delta}}
C_{\alpha\beta\gamma\delta}\,,
\ee
where the symmetrization over spinor indices denoted by the same
letter is implied. $F_{\alpha\beta}$,
$C_{\alpha\beta\gamma\delta}$ and their conjugated are totally
symmetric multispinors.

\abz Hodge duality for two-forms is translated as follows. By
definition, $P_{ij}$ and $*P_{ij}$ are related as
\be \label{hodge}
*P_{ij} = \frac{\sqrt{-g}}{2}\varepsilon_{ijkl}P^{kl},
\ee
where $g$ is the determinant of the metric and
$\varepsilon^{abcd}$ is Levi-Civita symbol ($\varepsilon^{0123}=
-\varepsilon_{0123}=1$). Then for spinor components we have
\be
*P_{\alpha\alpha} = -i P_{\alpha\alpha}, \qquad
*\bar{P}_{\dot\alpha\dot\alpha} = i
\bar{P}_{\dot\alpha\dot\alpha}\,.
\ee
In vector notation $P_{\alpha\alpha}, \bar
P_{\dot\alpha\dot\alpha}$ correspond to the (anti)self-dual  parts
$P^\pm_{ij}$ of antisymmetric tensor $P_{ij}$  defined by
\be \label{Fdual}
P^\pm_{ij}=\frac12 (P_{ij} \pm i*\!\!P_{ij}).
\ee

\abz $AdS_4$ spinor indices $A,B,\ldots$ unify left and right Weyl
spinor indices $A =(\alpha,\dot\alpha)$. These are raised and
lowered by the canonical $sp(4)$ form
\be
\gep_{AB} = \left(
\begin{array}{cc}
\gep_{\al\gb} & 0 \\
0 & \gep_{\dal\dgb} \\
\end{array}
\right)\,.
\ee
\renewcommand{\theequation}{B.\arabic{equation}}
\renewcommand{\thesubsection}{B}
\renewcommand{\thesection}{Appendix}
\makeatletter \@addtoreset{equation}{subsection} \makeatother
\subsection{Sketch of derivation of integrating flows} \label{App-B}

Let us sketch the idea of derivation of the equations
\eqref{M-flow} and \eqref{q-flow}. Since the method is similar for
the flows with respect to $\M$ and $\overline\M$ we confine
ourselves to $\p_\M$--flow with complex $\M$.

\abz Consider the most general decomposition of
$\p_\M\fV_{\al\dal}$ and $\p_\M\fh_{\al\dal}$ in the basis of
Kerr--Schild vectors \eqref{KS-vectors}
\be\label{gen-decomp}
\p_\M\fV_{\al\dal}=\sum_{I=1}^{4}t_I e_{I, \alpha\dot\alpha}\q
\p_\M\fh_{\al\dal}=\sum_{I,J=1}^{4}\Phi_{IJ}\hat e_{I,
\alpha\dot\alpha}\hat e_{J,
\gamma\dot\gamma}\fh^{\gamma\dot\gamma}
\ee with some set of functions $t_I(x,\M,\dots)$ and
$\Phi_{IJ}(x,\M,\dots)$. Applying $[\p_\M,d]$ to \eqref{dG} and
using $\p_\M\G=\p_\M\bar\G=0$ (see \eqref{dM0}) we obtain
\begin{eqnarray}
2\sum_I (-1)^{\sigma_I+\sigma_J} \Phi_{IJ} &=& t_J, \label{phi-t1}\\
2\sum_I (-1)^{\bar\sigma_I+\bar\sigma_J}\Phi_{IJ} &=&
t_J.\label{phi-t2}
\end{eqnarray}
Some components of $\Phi_{IJ}$ can be eliminated using the gauge
freedom of Cartan equations. Indeed, the gauge transformation of
vierbein is
\be\label{gauge}
\delta \fh_{\alpha\dot\alpha} = \D \xi_{\alpha\dot\alpha} +
\xi_{\alpha}{}^{\beta}\fh_{\beta\dot\alpha} +
\bar\xi_{\dot\alpha}{}^{\dot\beta}\fh_{\alpha\dot\beta}\,,
\ee
where $\xi_{\alpha\dot\alpha}, \xi_{\alpha\alpha},
\bar\xi_{\dot\alpha\dot\alpha}$ are arbitrary gauge parameters.
Using \eqref{gen-decomp} to extract pure gauge part in
\be
\gd_{\M}\fh_{\al\dal}=\sum_{I,J=1}^{4}\Phi_{IJ}\hat e_{I,
\al\dal}\hat\E_{J}\gd\M\,,
\ee
one can see that six parameters $\xi_{\al\al}$ and
$\bar\xi_{\dal\dal}$ make it possible to eliminate the
anti-symmetric part $\Phi_{[IJ]}$. Choose $\xi_{\al\dal}$-gauge
parameter in the  form
\be
\xi_{\al\dal}=\sum_{I=1}^{4}S_I\hat e_{I, \al\dal}\gd\M\,,
\ee
where $S_{I}(\G, \overline\G, \M, \dots)$ is some set of
functions. Using \eqref{FDA1}--\eqref{FDA2c} one obtains
\be
\D\xi_{\al\dal}=\sum_{I,J=1}^{4}(B_{[IJ]}+B_{(IJ)})\hat e_{I,
\al\dal}\hat\E_J\,,
\ee
where $B_{(IJ)}$ and $B_{[IJ]}$ are symmetric and antisymmetric.
Finally, making use of the gauge parameters $\xi_{\al\al}$ and
$\bar\xi_{\dal\dal}$ one eliminates the antisymmetric part
$\Phi_{[IJ]}$ . The symmetric part $\Phi_{(IJ)}$ is restricted by
\eqref{phi-t1}, \eqref{phi-t2} to the form
\be
\Phi_{(IJ)} = \left(\begin{array}{cccc}
\Phi_{11} & X & Z & Z \\
X & \Phi_{22} & Z & Z \\
Z & Z & \Phi_{33} & Y \\
Z & Z & Y & \Phi_{44}
\end{array}\right)\,.
\ee
Its off-diagonal part is parameterized by three parameters
$X,Y,Z$. The leftover gauge freedom in $\xi_{\alpha\dot\alpha}$
allows us to set $X,Y,Z$ to zero. At this stage, one gauge
parameter in $\xi_{\al\dal}$ remains free. An important
observation is the following. The gauge fixing that makes
$\Phi_{IJ}$ diagonal turns out to impose in addition to
\eqref{dM0} $\M$-independence condition on Maxwell two-form
\be
\p_\M F= 0\,.
\ee

\abz Thus, the gauge fixing leads to the following structure
functions
\be
\Phi_{IJ}=\frac12\gd_{IJ}\phi_J\q t_I=\phi_I\,.
\ee
The condition $[\p_\M, d]=0$ applied to
\eqref{FDA1}--\eqref{FDA2c} after somewhat  annoying but
straightforward calculation using the relations
\eqref{FDA1}--\eqref{FDA2c} and \eqref{DeA} gives the following
simple compatibility conditions
\be\label{eq1}
\f{\p\phi_1}{\p y}+(\G-\bar\G)\phi_1=0\q \f{\p\phi_3}{\p
r}+(\G+\bar\G)\phi_3=0\,,
\ee
\be\label{eq2}
\f{\p\phi_2}{\p y}+(\G-\bar\G)\phi_2=0\q \f{\p\phi_4}{\p
r}+(\G+\bar\G)\phi_4=0\,
\ee
along with the following constraints
\begin{eqnarray}
\phi_1+\phi_2+\phi_3+\phi_4&=&\f12\G+\f12\p_\M\mathcal{I}_1, \label{cons-1}\\
\phi_1+\phi_2-\phi_3-\phi_4&=&\f12\bar\G +
\frac{\G^2+\bar\G^2}{4\G\bar\G}\p_\M\mathcal{I}_1
 +\f{\G\bar\G}{4}\p_\M\mathcal{I}_2\,,\label{cons-2}
\end{eqnarray}
which can be equivalently rewritten as
\begin{eqnarray}
\phi_1+\phi_2&=&\frac{\G+\bar\G}{4}   (1+ r\p_\M\mathcal{I}_1 + \frac{1}{4r}\p_\M\mathcal{I}_2), \\
\phi_3+\phi_4&=&\frac{\G-\bar\G}{4}   (1- iy\p_\M\mathcal{I}_1 +
\frac{i}{4y}\p_\M\mathcal{I}_2).
\end{eqnarray}
Let us note, that the conditions \eqref{cons-1}, \eqref{cons-2}
arise if $\p_\M$-flow is applied to \eqref{FI1} and \eqref{FI2},
respectively.

\abz Solutions for $\phi_I$ exist for any values of
$\mathcal{I}_1$ and $\mathcal{I}_2$. However, the case with
$\p_\M\mathcal{I}_{1,2}\neq0$, although being consistent with
\eqref{eq1}, \eqref{eq2} turns out to be incompatible with the
reality of the metric and even in the complex case it does not
seem to lead to any simplification. Therefore, we demand
\be\label{dI2}
\p_\M\mathcal{I}_1=0\q \p_\M\mathcal{I}_2=0\,.
\ee
Performing straightforward integration of \eqref{eq1} and
\eqref{eq2} we obtain \eqref{1}, \eqref{2} and \eqref{cnstr-1}
with $\al_{1,2}(r)$ and $\gb_{1,2}(y)$ arising as the integration
parameters.

\abz The analysis for $\p_\e$--flow is analogous leading to the
same differential equations \eqref{eq1}, \eqref{eq2} for
$\psi_I$--functions along with the constraints
\begin{eqnarray}
\psi_1+\psi_2+\psi_3+\psi_4&=& -\f{\G\bar\G}{2} +   \frac12\p_\e\mathcal{I}_1\\
\psi_1+\psi_2-\psi_3-\psi_4&=&
\frac{\G^2+\bar\G^2}{4\G\bar\G}\p_\e\mathcal{I}_1 +
\f{\G\bar\G}{4}\p_\e\mathcal{I}_2
\end{eqnarray}
or, equivalently,
\begin{eqnarray}
\psi_1+\psi_2&=&-\frac{\G\bar\G}{4}   (1- 2r^2\p_\e\mathcal{I}_1 - \frac12\p_\e\mathcal{I}_2), \\
\psi_3+\psi_4&=&-\frac{\G\bar\G}{4}   (1- 2y^2\p_\e\mathcal{I}_1 +
\frac12\p_\e\mathcal{I}_2)\,.
\end{eqnarray}
It is convenient to set $\p_\e\mathcal{I}_1=0$ reproducing
\eqref{3}, \eqref{4} and \eqref{cnstr-2}, where
$\gga=\p_\e\mathcal{I}_2$. Finally, the condition $[\p_\M,
\p_\e]=0$ leads to \eqref{cnstr-3}. The reason to keep
$\p_\e\mathcal{I}_2\neq 0$ is that it allows to reduce black hole
metric to convenient Kerr--Schild form when $\M=\overline\M$.

\abz Finally, the case with real $\M=\overline\M$ can be
considered separately as it admits some simplification of the
metric. When $\M$ is real one is left with the only mass flow
instead of two for complex $\M$. This case provides the same
system of differential equations for the structure functions
\eqref{eq1}, \eqref{eq2}. Analogously, to have the complete set of
consistency conditions for $\phi_{I}$ one acts with $\p_\M$ on the
first integrals \eqref{FI1}, \eqref{FI2}. Note, that in this case
one should set $\M=\overline\M$ on the right hand sides of
\eqref{FI1}, \eqref{FI2} prior their differentiation. Pretty much
as in the complex case, it is convenient to demand \eqref{dI2}. As
a result, in addition to \eqref{eq1}, \eqref{eq2} one gets the
following constraints
\be\label{cons-real}
\phi_1+\phi_2+\phi_3+\phi_4=\f12(\G+\bar\G)\,,\quad
\phi_1+\phi_2-\phi_3-\phi_4=\f12(\G+\bar\G)\,.
\ee
General solution of \eqref{eq1}, \eqref{eq2}, \eqref{cons-real} is
\be\label{1-real}
\phi_{1}=\f{G+\bar\G}{2}\al_1(r)\q
\phi_{2}=\f{G+\bar\G}{2}\al_2(r)\,,
\ee
\be
\phi_{3}=\f{G-\bar\G}{2}\gb_1(y)\q
\phi_{4}=\f{G-\bar\G}{2}\gb_2(y)\,,
\ee
where
\be
\al_1(r)+\al_2(r)=1\q \gb_1(y)+\gb_2(y)=0\,.
\ee
Note, that \eqref{1-real} differs from \eqref{1} by 2 factor and
the constraints for $\gb_{1,2}$ differ from that in
\eqref{cnstr-1}. This integration allows one to fix gauge
parameters $\al_2=\gb_1=\gb_2=0$ to get \eqref{gauge-1}.

\abz Let us note, that for each of the integrating flows
$\p_{\chi}=(\p_\M$, $\p_{\overline{\M}}$, $\p_{\e})$ the
integrability condition
\be\label{d-cons}
[\p_\chi, d]=0
\ee
provides the same differential equations for the flow structure
functions \eqref{eq1}, \eqref{eq2}. These equations are not
sufficient for the consistency \eqref{d-cons}. The rest of the
conditions, such as \eqref{cons-1}, \eqref{cons-2} result from the
requirement that \eqref{FI1}, \eqref{FI2} are constant in BHUS.
Together with \eqref{eq1} and \eqref{eq2} they satisfy
\eqref{d-cons}. Having solved \eqref{d-cons} one is left with
\be
[\p_{\chi}, \p_{\chi'}]=0
\ee
which is straightforward to analyze.

\renewcommand{\theequation}{C.\arabic{equation}}
\renewcommand{\thesubsection}{C}
\renewcommand{\thesection}{Appendix}
\makeatletter \@addtoreset{equation}{subsection} \makeatother

\subsection{Integration of Kerr--Schild vectors} \label{App-C}

Consider the general case of complex $\M$. To restore Kerr--Schild
vectors in terms of their $AdS_4$ counterparts \eqref{e-vec1},
\eqref{e-vec2} via the flow integration we differentiate
\eqref{KS-vectors} along the flows and use their definition to
obtain
\begin{eqnarray}
\p_\M \hat k_{\alpha\dot\alpha} &=& -\frac{\alpha_2
r}{\hat\Delta_r} \hat k_{\alpha\dot\alpha}, \quad
\p_\M\hat l^{+-}_{\alpha\dot\alpha} = -\frac{\gb_2 y}{i\hat\Delta_y} \hat l^{+-}_{\alpha\dot\alpha},\\
\p_\M \hat n_{\alpha\dot\alpha} &=& -\frac{\alpha_1
r}{\hat\Delta_r} \hat n_{\alpha\dot\alpha},  \quad \p_\M \hat
l^{-+}_{\alpha\dot\alpha} = -\frac{\gb_1 y}{i\hat\Delta_y} \hat
l^{-+}_{\alpha\dot\alpha}\,.
\end{eqnarray}
Analogously, applying $\p_\e$--flow we obtain
\begin{eqnarray}
\p_\e \hat k_{\alpha\dot\alpha} &=& \frac{\theta_2}{2\hat\Delta_r}
\hat k_{\alpha\dot\alpha}, \quad
\p_\e \hat l^{+-}_{\alpha\dot\alpha} = -\frac{\vartheta_2}{2\hat\Delta_y} \hat l^{+-}_{\alpha\dot\alpha},\\
\p_\e \hat n_{\alpha\dot\alpha} &=& \frac{\theta_1}{2\hat\Delta_r}
\hat n_{\alpha\dot\alpha},  \quad \p_\e \hat
l^{-+}_{\alpha\dot\alpha} = -\frac{\vartheta_1}{2\hat\Delta_y}
\hat e_{4,\alpha\dot\alpha}.
\end{eqnarray}
Recall, that the parameters $\al_{1,2}(r), \gb_{1,2}(y),
\theta_{1,2}(r), \vartheta_{1,2}(y)$ arise as the integration
constants in \eqref{eq1}, \eqref{eq2}. Integration with the
constraint \eqref{func-fix} gives \eqref{e-vec1}, \eqref{e-vec2}.

\abz To obtain \eqref{e-form1}, \eqref{e-form2} we contract the
second equation of \eqref{M-flow} with all Kerr--Schild vectors
\eqref{KS-vectors}. This gives
\begin{eqnarray}
\p_\M \hat K &=&\frac{\alpha_2 r}{\hat\Delta_r} ( \hat N-\hat K),
\quad
\p_\M \hat L^{+-} =\frac{\gb_2 y}{i\hat\Delta_y} (\hat L^{-+}-\hat L^{+-}), \label{E1-flow} \\
\p_\M \hat N &=& \frac{\alpha_1 r}{\hat\Delta_r} (\hat K-\hat N),
\quad \p_\M \hat L^{-+} =\frac{\gb_1 y}{i\hat\Delta_y} (\hat
L^{+-}-\hat L^{-+}). \label{E2-flow}
\end{eqnarray}
The analysis for $\p_\e$--flow is analogous and the integration at
the condition \eqref{func-fix} gives \eqref{e-form1},
\eqref{e-form2}.

\renewcommand{\theequation}{D.\arabic{equation}}
\renewcommand{\thesubsection}{D}
\renewcommand{\thesection}{Appendix}
\makeatletter \@addtoreset{equation}{subsection} \makeatother

\subsection{Vector form of $AdS_4$ unfolded system} \label{App-D}

$AdS_4$ unfolded system in vector notation reads as
\begin{eqnarray}
D_i V_j &=& \kappa_{ij},  \label{vda-1}\\
D_k\kappa_{ij} &=& \lambda^2 (g_{kj}V_i - g_{ki}V_j).
\label{vda-2}
\end{eqnarray}
Let us introduce antisymmetric tensor $F_{ij} = F^+_{ij}+
F^-_{ij}$  by
\be
\kappa^+_{ij} = \rho F^+_{ij}, \quad \kappa^-_{ij} = \bar\rho
F^-_{ij},
\ee
where
\be
\rho= -\lambda^2\G^{-3}, \quad \bar\rho= -\lambda^2\bar\G^{-3},
\ee
and
\be
F^+_{ij}F^{+ij} = -\G^4, \quad F^-_{ij}F^{-ij} = -\bar\G^4,
\ee
where $\pm$ denote (anti)self-dual parts of the corresponding
two-forms. One can easily check that $F_{ij}$ fulfills Maxwell
equation and Bianchi identities
\be
D_k F^k{}_i = 0, \quad D_{[i}F_{jk]} = 0.
\ee
Then the system \eqref{vda-1}, \eqref{vda-2} can be equivalently
rewritten as
\begin{eqnarray}
D_i V_j &=& \rho F^+_{ij} + \bar\rho F^-_{ij}, \label{v-fda1}\\
D_k F^+_{ij} &=& V^p C^+_{pkij}, \label{v-fda2}\\
D_k F^-_{ij} &=& V^p C^-_{pkij},\label{v-fda3}
\end{eqnarray}
where $C^\pm_{ijkl}$ are the following quadratic combinations of
$F^\pm_{ij}$
\be \label{CFdualp}
C^{+}_{ijkl} = -\G^{-1} \left( 2F^{+}_{ij}F^{+}_{kl}
+F^{+}_{ik}F^{+}_{jl}-F^{+}_{il}F^{+}_{jk}+ \frac{\G^4}{4}\left(
g_{ik}g_{jl} - g_{il}g_{jk} \right) \right),
\ee
\be \label{CFdualm}
C^{-}_{ijkl} = -\bar\G^{-1} \left( 2F^{-}_{ij}F^{-}_{kl}
+F^{-}_{ik}F^{-}_{jl}-F^{-}_{il}F^{-}_{jk}+
\frac{\bar\G^4}{4}\left( g_{ik}g_{jl} - g_{il}g_{jk}  \right)
\right).
\ee
Note that $C^\pm_{ijkl}$ describe the (anti)self-dual parts of the
black hole Weyl tensor.

\abz The Killing projectors in vector indices are
\begin{eqnarray}
\Pi^\pm_{ij} = \frac12 g_{ij} \pm \G^{-2} F^+_{ij}, \\
\bar\Pi^\pm_{ij} = \frac12 g_{ij} \pm \bar\G^{-2} F^-_{ij}.
\end{eqnarray}
They possess the following obvious properties
 \be
 \Pi^\pm_{i}{}^{j}  \Pi^\pm_{jk} = \Pi^\pm_{ik}, \quad  \Pi^\pm_{i}{}^{j}  \Pi^\mp_{jk} = 0,
 \ee
and
 \be
\bar \Pi^\pm_{i}{}^{j}  \bar\Pi^\pm_{jk} = \bar\Pi^\pm_{ik}, \quad
\bar\Pi^\pm_{i}{}^{j}  \bar\Pi^\mp_{jk} = 0.
 \ee
In spinor notation these projectors yield  \eqref{introP}.

\abz The Mutually orthogonal null-vectors
\eqref{kn-def}--\eqref{ll-def} are the following projections of a
Killing vector
\begin{eqnarray}
k_i &=& \frac{1}{(V^{+}V^-)}V^{-}_i, \qquad
n_i = \frac{1}{(V^{+}V^-)}V^{+}_i,  \label{kn-vec} \\
l^{+-}_i &=& \frac{1}{(V^{+-}V^{-+})}V^{+-}_i, \qquad l^{-+}_i =
\frac{1}{(V^{+-}V^{-+})}V^{-+}_i, \label{ll-vec}
\end{eqnarray}
where $(AB)=A_iB^i$ and
\begin{eqnarray}
V^{-}_i &=& \Pi^-_i{}^j \bar\Pi^-_{jk} V^k, \qquad
V^{+}_i = \Pi^+_i{}^j \bar\Pi^+_{jk} V^k, \\
V^{+-}_i &=& \Pi^+_i{}^j \bar\Pi^-_{jk} V^k \qquad V^{-+}_i =
\Pi^-_i{}^j \bar\Pi^+_{jk} V^k.
\end{eqnarray}
Let us also note that
\be
\Pi_i{}^j \bar\Pi_{jk}  = \bar\Pi_i{}^j \Pi_{jk}.
\ee
The vectors  \eqref{kn-vec}, \eqref{ll-vec} define the four
geodesic congruences
\be
k^jD_j k_i = 0, \quad n^jD_j n_i = 0, \quad l^{+-j}D_j l^{+-}_i =
0, \quad l^{-+j}D_j l^{-+}_i = 0.
\ee
Now one can check that consistency of the Killing-Maxwell system
\eqref{v-fda1}--\eqref{v-fda2} demands the function $\rho$ to be
of the form \eqref{rho-sol}. The Riemann tensor reads as
\begin{align}
R_{ijkl} =& \lambda^2(g_{ik}g_{jl}-g_{il}g_{jk}) +2(e^2+\g^2)(g_{ik}T_{jl}+g_{jl}T_{ik}- g_{il}T_{jk}-g_{jk}T_{il})\notag \\
&+ 6(\M-2(e^2+\g^2)\bar\G)C^+_{ijkl} +
6(\overline\M-2(e^2+\g^2)\G)C^-_{ijkl},
\end{align}
where the energy-momentum tensor $T_{ij}$ has the following simple
form
\be
T_{ij} = 2 F^+_{ik}F^-_j{}^k = 2 F^-_{ik}F^+_j{}^k.
\ee
Rewriting \eqref{fege} and \eqref{fegec} in vector notation yields
\begin{eqnarray}
F_{ij}k^j &=& -\frac{\G^2 + \bar\G^2}{2} k_i , \quad F_{ij}n^j = \frac{\G^2 + \bar\G^2}{2} n_i, \\
F_{ij}l^{+-j}&=& \frac{\G^2 - \bar\G^2}{2} l^{+-}_i ,\quad
F_{ij}l^{-+j} = -\frac{\G^2 -\bar\G^2}{2} l^{-+}_i
\end{eqnarray}
and
\begin{eqnarray}
*\!F_{ij}k^j &=& \frac{i(\G^2 - \bar\G^2)}{2} k_i , \quad *\!F_{ij}n^j = \frac{i(\G^2 - \bar\G^2)}{2} n_i, \\
*\!F_{ij}l^{+-j}&=& -\frac{i(\G^2 + \bar\G^2)}{2} l^{+-}_i ,\quad
*\!F_{ij}l^{-+j} = \frac{i(\G^2 +\bar\G^2)}{2} l^{-+}_i.
\end{eqnarray}

\renewcommand{\theequation}{E.\arabic{equation}}
\renewcommand{\thesubsection}{E}
\renewcommand{\thesection}{Appendix}
\makeatletter \@addtoreset{equation}{subsection} \makeatother
\subsection{Some useful unfolded system properties} \label{App-E}

Considered unfolded systems have a number of important properties,
such as the existence of Killing--Yano tensors and additional
Killing vector. Let us show this in some detail for the case of
$AdS_4$.

\abz Using \eqref{FDA2-ads} and \eqref{F-vac} we find
\be \label{V_F/G}
D\left(\frac{1}{\G^3}F_{\alpha\alpha}\right) =
-h_{\alpha}{}^{\dot\alpha}V_{\alpha\dot\alpha}.
\ee
Therefore, the Maxwell tensor generates Killing--Yano tensor
\be \label{yka}
Y_{\al\al}=\f{i}{\G^3}F_{\al\al}, \qquad
\bar{Y}_{\dot\al\dot\al}=-\f{i}{\bar{\G}^3}\bar{F}_{\dot\al\dot\al}.
\ee
Indeed, \eqref{yka} gives
\be \label{yano}
D_{\al\dal}Y_{\al\al}=0\,,\quad
D_{\gb\dal}Y^{\gb}{}_{\al}+D_{\al\dgb}\bar{Y}^{\dgb}{}_{\dal}=0\,,
\ee
which is equivalent  to the Killing--Yano equation in vector
indices \cite{Yano}
\be
D_{(k}Y_{m)n}=0, \quad Y_{mn}=-Y_{nm}.
\ee
The next observation is that the Hodge dual tensor (see Appendix
\ref{App-A}) $*Y_{ij}$ that has the following irreducible
components
\be \label{cky}
*Y_{\al\al}=\f{1}{\G^3}F_{\al\al}, \qquad
*\bar{Y}_{\dot\al\dot\al}=\f{1}{\bar{\G}^3}\bar{F}_{\dot\al\dot\al}.
\ee
is the closed Killing--Yano tensor since it fulfils the
equation\footnote{Note that covariant differential acts on forms
as ordinary de Rham differential.}  $d*Y = 0$, {\it i.e}.,
\be
\partial_{[i}*\!Y_{jk]}=0, \quad *Y_{mn}=-*\!Y_{nm}.
\ee
where brackets denote antisymmetrization over indices. It is
obvious that
\be
*Y_{mn}=-\frac{1}{\lambda^2}\kappa_{mn},
\ee
where $\kappa_{mn}$ is given by \eqref{k2f}.

\abz One can see that it is possible to express Killing vector
$V_{\alpha\dot\alpha}$ as
\be
V^i = \frac{1}{3}D_j *\!Y^{ji}.
\ee
Note that  another Killing vector can be constructed from $V^i$ by
means of the second-rank Killing tensor $K_{ij} $ generated by the
Killing--Yano tensor $Y_{ij}$ (see \cite{Carter3})
\be
\phi_i = K_{ij}V^j, \qquad K_{ij} = Y_{ik}Y^{k}{}_j.
\ee
In spinor notation this gives the following relation between
Killing vectors
\be \label{phi-kill}
\phi_{\alpha\dot\alpha} = \frac{1}{4\G^3\bar\G^3}
F_{\alpha}{}^{\beta} \bar{F}_{\dot\alpha}{}^{\dot\beta}
V_{\beta\dot\beta} - \frac14\left(\frac{1}{\G^2} +
\frac{1}{\bar\G^2}\right) V_{\alpha\dot\alpha}.
\ee
One can make sure that it solves the Killing equation
\be \label{phi-kill-eq}
D\phi_{\alpha\dot\alpha} = \frac12 h_{\alpha}{}^{\dot\beta}
\bar\varphi_{\dot\alpha\dot\beta} +\frac12
h^{\beta}{}_{\dot\alpha} \varphi_{\alpha\beta},
\ee
where $\varphi_{\alpha\alpha}, \bar\varphi_{\dot\alpha\dot\alpha}$
are the (anti)self-dual components of the Killing two-form
\be\label{phi-kill-2f}
\varphi_{\alpha\alpha} = -\frac{1}{2\G^3}
\left(I_1+\frac{\lambda^2}{\bar\G^2}\right) F_{\alpha\alpha}
-\frac{1}{2\bar\G^3}V_{\alpha}{}^{\dot\alpha}V_{\alpha}{}^{\dot\alpha}\bar{F}_{\dot\alpha\dot\alpha}.
\ee
Note, that when $C_4 = C_2^2$ this Killing vanishes
$\phi_{\alpha\dot\alpha} = 0$ and $\varphi_{\alpha\alpha} = 0$.

\abz As a result the global symmetry parameter $K_{AB}$
\eqref{ads-kappa} generates another global symmetry parameter
$\tilde{K}_{AB}$ in the following way
\be \label{ads-phi} \tilde{K}_{AB} = \left(
\begin{array}{cc}
\varphi_{\alpha\beta} & \lambda \phi_{\alpha\dot\beta} \\
\lambda \phi_{\beta\dot\alpha} & \bar\varphi_{\dot\alpha\dot\beta}
\end{array}
\right), \quad D_0\tilde{K}_{AB}  = 0\,.
\ee
Its $sp(4)$  invariants read
\begin{eqnarray}
\mathcal{C}_2 &=& \tilde{K}_{AB}\tilde{K}^{AB}  = I_1 I_2,\\
\mathcal{C}_4 &=& \text{Tr}(\tilde{K}^4) =  \frac{I_2^2}{4} (I_1^2
+ \lambda^2 I_2)\,.
\end{eqnarray}

\abz Remarkably, the existence of the Killing--Yano tensor and additional
Killing vector also takes place in the black hole unfolded system
\eqref{FDA1}--\eqref{FDAh}. One can show that the formulae
\eqref{V_F/G}--\eqref{phi-kill} are valid in BHUS upon
redefinition
\be
D\to\fD\q V_{\al\dal}\to\fV_{\al\dal}\,.
\ee

\renewcommand{\theequation}{F.\arabic{equation}}
\renewcommand{\thesubsection}{F}
\renewcommand{\thesection}{Appendix}
\makeatletter \@addtoreset{equation}{subsection} \makeatother

\subsection{Comment on Plebanski--Demianski solution} \label{App-F}

The considered unfolded system \eqref{FDA1}--\eqref{FDA2c} has
been shown to describe generic Carter--Plebanski family of
metrics. This family can be obtained by some limiting procedure
from the so called Plebanski--Demianski metric \cite{PLB}, which
is believed to be the most general D-type solution of vacuum
Einstein--Maxwell equations with aligned principle directions.
Although it has the same number of parameters as the
Carter--Plebanski solution all of them are continuous unlike those
in Carter--Plebanski case with one being discrete. Physical
meaning of this additional continuous parameter according to
\cite{PLB} is acceleration. The metric generalizes
Carter--Plebanski solution reproducing the latter in
non-accelerated limit. In this Appendix we demonstrate that the
constructed black hole unfolded system does not contain
Plebanski--Demianski metric.

\abz Consider Plebanski--Demianski class of solutions of the
Einstein--Maxwell equations with  electric and magnetic charges
and the cosmological constant. Up to notation, its original form
reads as
\be \label{PDm}
g = \frac{1}{(1- q
p)^2}\left(\frac{Q(q)(d\tau-p^2d\psi)^2}{q^2+p^2} -
\frac{P(p)(d\tau+q^2d\psi)^2}{q^2+p^2} - \frac{q^2+p^2}{Q(q)}dq^2
- \frac{q^2+p^2}{P(p)}dp^2\right)\,,
\ee
where
\begin{eqnarray}
Q(q) &=&  (\lambda^2/2 + \gamma  + e^2) -2 m q + \varepsilon q^2 - 2 nq^3 -(\lambda^2/2- \gamma +\g^2)q^4, \\
P(p) &=& (\lambda^2/2 + \gamma -\g^2) +2 n p - \varepsilon p^2 + 2
mp^3 - (\lambda^2/2- \gamma -e^2 )p^4.
\end{eqnarray}
Choosing the vector potential in the form
\be
A = \frac{q}{q^2+p^2}d\tau - \frac{q p^2}{q^2+p^2}d\psi
\ee
one can check that the following equations hold
\be
R_{ij} = 3\lambda^2 g_{ij} - 2(e^2+\g^2)(F_{ik}F_{j}{}^k - \frac14
g_{ij} F^{kl} F_{kl}),
\ee
where $F_{ij}$ is the Maxwell tensor $F=dA$
\be \label{F-PD}
F = \frac{1}{(q^2 + p^2)^2} \left[  (d\tau -p^2 d\psi)\wedge
(q^2-p^2) dq  + 2(d\tau +q^2 d\psi)\wedge pq dp \right]\,.
\ee
The Weyl tensor of \eqref{PDm} is built of \eqref{F-PD} according
to \eqref{weyl} thus having a chance to be described by
\eqref{FDA1}--\eqref{FDA2c} with some Killing vector $V^i$. Let us
show that this is not the case for any real $V^i$.

\abz Assuming, that Plebanski--Demianski metric can be described
by BHUS we use its Maxwell tensor \eqref{F-PD} to express
corresponding Killing vector $V^i$ via \eqref{dG} in two different
ways
\begin{eqnarray}
(a)& \quad &V^i = \frac{1}{2\G^4} h_{\alpha\dot\alpha}{}^i F^{\alpha\gamma}\partial_{\gamma}{}^{\dot\alpha} \G, \\
(b)& \quad &V^i = \frac{1}{2\bar\G^4} h_{\alpha\dot\alpha}{}^i
\bar{F}^{\dot\alpha\dot\gamma}\partial^{\alpha}{}_{\dot\gamma}
\bar\G\,.
\end{eqnarray}
By construction, in the Carter--Plebanski case these two formulae
give the same result with real $V^{i}$. For \eqref{F-PD} this is
not the case and we obtain two complex-conjugated Killing vectors
\be
V_{(a)}^i = \{1,-i,0,0\}, \quad V_{(b)}^i = \{1,i,0,0\}.
\ee
Thus, Plebanski--Demianski metric \eqref{PDm} is governed by a
complex Killing vector or, equivalently, two real ones. Its
unfolded system requires some modification to include complex
Killing vector and will be considered elsewhere.


\vspace{1cm}


\begin{thebibliography}{36}

\bibitem{DMV}
V.E. Didenko, A.S. Matveev and M.A. Vasiliev, Phys. Lett. {\bf
B665} (2008) 284

\bibitem{review}
D. Kramer, H. Stephani, E. Herlt and M.A.H. MacCallum, {\it Exact
solutions of Einstein's field equations}, Cambridge University
Press, Cambridge,  1980

\bibitem{Car}
B. Carter, Commun. Math. Phys. {\bf 10} (1968) 280

\bibitem{Carter4}
B. Carter,  Phys. Rev. {\bf 174} (1968) 1559

\bibitem{Pleb}
J.F. Plebanski, Ann. Phys. (NY) {\bf 90} (1975) 196

\bibitem{PV}
S.F. Prokushkin and M.A. Vasiliev, Nucl. Phys. {\bf B545} (1999)
385

\bibitem{Gol}
M.A. Vasiliev, \emph{Higher spin gauge theories: Star-product and
$AdS$ space}, in M.A. Shifman ed., The many faces of the
superworld (World Scientific, 2000), [{\tt hep-th/9910096}]

\bibitem{Ann}
M.A. Vasiliev, Ann. Phys. (NY) {\bf 190} (1989) 59

\bibitem{more}
M.A. Vasiliev, Phys. Lett. {\bf B285} (1992) 225

\bibitem{QG}
M.A. Vasiliev, Int. J. Mod. Phys. {\bf D5} (1996) 763-797

\bibitem{sor}
D. Sorokin, \emph{Introduction to the classical theory of higher
spins}, [{\tt hep-th/0405069}]

\bibitem{SSS}
A.~Sagnotti, E.~Sezgin and P.~Sundell, ``On Higher Spins with a
Strong $Sp(2,{\mathbb R})$ Condition,'' Proceedings of the First
Solvay Workshop on Higher-Spin Gauge Theories (Brussels, May
2004), [{\tt hep-th/0501156}]

\bibitem{solv}
X. Bekaert, S. Cnockaert, C. Iazeolla and M.A. Vasiliev,
\emph{Nonlinear higher spin theories in various dimensions},
Proceedings of the First Solvay Workshop on Higher-Spin Gauge
Theories (Brussels, May 2004), [{\tt hep-th/0503128}]

\bibitem{Vas}
M.A. Vasiliev, Nucl.Phys. {\bf B616} (2001) 106-162; Erratum-ibid.
{\bf  B652} (2003) 407

\bibitem{Newman}
E.T. Newman and  A.I. Janis,  J. Math. Phys. {\bf 6} (1965) 915;
E.T. Newman, E. Couch, R. Chinnapared, A. Exton, A. Prakash and R.
Torrence, J. Math. Phys. {\bf 6} (1965) 918

\bibitem{Petrov}
A.Z. Petrov,  \emph{The classification of spaces defining
gravitational fields}, Scientific Proceedings of Kazan State
University  {\bf 114} (1954) 55

\bibitem{K-S}
R.P. Kerr and A. Schild, Proc. Symp. Appl. Math. {\bf 17} (1965)
199

\bibitem{Carter3}
B. Carter,  J. Math. Phys. {\bf 28} (1987) 1535

\bibitem{Papa}
A. Papapetrou, Ann. Inst. H. Poincar\'e {\bf A4} (1966) 83

\bibitem{BL}
R.H. Boyer and R.W. Lindquist,  J. Math. Phys. {\bf 8} (1967) 265

\bibitem{Carter2}
B. Carter,  Commun.  Math.  Phys. {\bf 17}  (1970)  233


\bibitem{Gibon2}
Z.-W. Chong, G.W. Gibbons, H. Lu and C.N. Pope, Phys. Lett. {\bf
B609} (2005) 124-132

\bibitem{Gib}
G.W. Gibbons, H. Lu, Don N. Page and C.N. Pope, J. Geom. Phys.
{\bf 53}  (2005) 49-73

\bibitem{SV}
O.V. Shaynkman and M.A. Vasiliev, Theor.\ Math.\ Phys.\  {\bf 123}
(2000) 683 [Teor.\ Mat.\ Fiz.\ {\bf 123} (2000) 323]

\bibitem{Reall}
R. Emparan, H.S. Real, Phys. Rev. Lett. {\bf 88} (2002) 101101

\bibitem{Pravdova}
V. Pravda and A. Pravdova, Gen. Rel. Grav {\bf 37} (2005)
1277-1287

\bibitem{Frol}
V. Frolov and D. Kubiznak, Phys. Rev. Lett. {\bf 98} (2007) 011101

\bibitem{PRD}
M.A. Vasiliev, Phys. Rev. {\bf D66} (2002) 066006

\bibitem{Vas-Sp}
M.A. Vasiliev,  {\it Relativity, Causality, Locality, Quantization
and Duality in the $Sp(2M)$ Invariant Generalized Space-Time},
Contribution to the Marinov's Memorial Volume, M.Olshanetsky and
A.Vainshtein Eds, World Scientific

\bibitem{Plyush}
M. Plyushchay, D. Sorokin and M. Tsulaia,  {\it $GL$ flatness of
$OSp(1|2n)$ and higher spin field theory from dynamics in
tensorial spaces}, [{\tt hep-th/0310297}]

\bibitem{DV}
V. Didenko and M.A. Vasiliev, J. Math. Phys. {\bf 45} (2004)
197-215


\bibitem{Yano}
K. Yano, Ann. Math. {\bf 55} (1952) 328

\bibitem{Frolov2}
V.P. Frolov and D. Kubiznak, Class. Quant. Grav. {\bf 25} (2008)
154005

\bibitem{PLB}
J.F. Pleba\'nski and M. Demia\'nski,  Ann. Phys. (NY)  {\bf 98}
(1976) 98-127


\end{thebibliography}
\end{document}